\documentclass[aps,prd,preprint,groupedaddress,nofootinbib,showkeys]{revtex4-1}
\usepackage{amsmath}
\usepackage{amssymb}
\usepackage{graphicx}
\usepackage{lineno}
\usepackage{color}
\usepackage{comment}
\newcommand{\xt}{(\mathbf{x},t)}

\newcommand{\deltav}{\delta^3(\mathbf{x}' - \mathbf{x}_1 -\mathbf{v}(t'-t_1) ) }
\newcommand{\thetaprim}[1]{\Theta(t' - t_{#1})}
\newcommand{\rect}{\thetaprim{1}-\thetaprim{2}}

\newcommand{\xw}{(\mathbf{x},\omega)}
\newcommand{\kw}{(k_x,k_y,z,\omega)}

\begin{document}

%\preprint{}

\title{Influence of a planar boundary on the electric field emitted by a particle shower}


\author{ Daniel Garc\'\i a-Fern\'{a}ndez$^1$, Beno\^\i t Revenu$^{1,2}$, 
             Antony Escudie$^1$, Lilian Martin$^{1,2}$ \\
            $^1\,$Subatech, Institut Mines-T\'{e}l\'{e}com Atlantique, CNRS, Universit\'{e} de Nantes, Nantes, France\\
            $^2\,$Unit\'{e} Scientifique de Nan\c{c}ay, Observatoire de Paris, CNRS, PSL, UO/OSUC, Nan\c{c}ay, France }
%\email{daniel.garcia-fernandez@subatech.in2p3.fr}
%\thanks{}


\date{\today}

\begin{abstract}
The radio detection of cosmic rays consists in the estimation of the properties of a primary cosmic ray
by observing the electric field emitted by the extensive air shower (EAS) created when the primary
cosmic ray enters the atmosphere. This technique is fully operative nowadays and presents a good
degree of maturity. In addition, several projects intend to employ this technique for the detection
of neutrinos. In order for the technique to be useful, accurate methods for computing the electric
field created by a particle shower in the context of a particular experiment must exist. 
Although current ground-based radio experiments lie on the air-soil interface and some
planned experiments on the South Pole envision antennas near the air-ice interface,
most of the analytical approaches and Monte Carlo codes used for
calculating the electric field either do not take into account the effect of the boundary
or calculate the radiation fields only (direct, reflected and transmitted radiation fields).
When the particle shower and the antenna are close to the boundary, compared to the
observation wavelength, the far-field approximation breaks down, which is the case for
the low-frequency EXTASIS experiment, for instance. We present in this work a new formula
for calculating the exact field emitted by a particle track in two semi-infinite media separated
by a planar boundary. We also explore the validity of the far-field approximation and make
some predictions for EAS using a simple shower model.
\end{abstract}

\pacs{}
\keywords{high energy cosmic rays, neutrinos, extensive air showers, dense media showers, radio detection, surface wave}

\maketitle

\section{Introduction}

Due to the low flux of cosmic rays with energies above $10^{15}$ eV
($\sim 3\cdot 10^{-4}$~m$^{-2}$~min$^{-1}$), high-energy cosmic
rays cannot be directly detected and indirect methods must be used. 
When a cosmic ray enters the atmosphere, it creates an
extensive air shower (EAS) that can be detected and whose properties give us information on
the primary cosmic ray. The detection of these EAS is carried out nowadays with the help of
three main techniques: the surface detection technique, consisting in the measurement of the
particles arriving at ground level and, the fluroescence 
detection, which consists in the collection of the fluorescent light left by the EAS upon its
passage through the atmosphere, and finally, the radio detection technique.

The charged particles in the EAS emit an electric field that can be detected using antennas.
This is the principle of the radio detection technique. The radio technique is well established
nowadays \cite{frankreview}, and it presents the capability of measuring the most important
properties of cosmic rays, namely, their arrival direction, energy, and composition with urcentainties
competitive with respect to the fluorescence detection \cite{lofarnature,gdas}. Radio experiments
such as CODALEMA \cite{icrcarxiv}, AERA \cite{aera}, LOFAR \cite{lofar} or Tunka-Rex \cite{tunka}
detect and analyze cosmic rays on a routinary basis, proving the maturity of the technique
at the present time.

Radio detection could be used for detecting neutrinos as well, since
neutrinos create a particle shower when they interact within dense media. Initiatives like
ARIANNA \cite{arianna} and ARA \cite{ara} try to pave the way for the radio detection of
neutrinos on the South Pole. Due to the low cost of the detectors, and if radio detection
of neutrinos is feasible, it could be possible to
build a radio-based neutrino observatory with an instrumentation volume bigger than the
%current neutrino observatories (for instance, IceCube \cite{icecube}) 
planned new-generation neutrino observatories (IceCube-Gen2 \cite{icecubegen2},
KM3NeT 2.0 \cite{km3net}, or GRAND \cite{grandwhite}) with a relatively low cost.

In order for the radio technique to be useful, the electric field emitted by the particle showers
must be understood. Monte Carlo codes such as SELFAS3 \cite{selfas,sdp}, ZHAireS \cite{zhaires}
or CoREAS \cite{coreas} allow the calculation of the emitted field, which enables the reconstruction
of the primary cosmic ray upon comparison with the measured data.

Although ZHAireS can take into account the reflection on the ground for calculating the field
received by balloon-borne antennas \cite{reflected}, when calculating the field
seen by antennas near ground level
the effect of the boundary is usually not taken into account by these codes. 
Only direct fields, as if there were no boundary, are calculated.
This is justified by the fact that if the
distance between the emitting particles and the receiving antenna is large with respect to the
observation wavelength (what is called the far field or radiation field), 
the influence of the boundary can be included either by
calculating the reflected field (applying the Fresnel coefficients on the direct field) 
or knowing the reception pattern
of the antenna when the boundary is present, by means of the reciprocity theorem \cite{balanis,antennas}.
However, this is not true if the observation wavelength is large enough 
(or, equivalently, the frequency low enough), 
as such is the case for the EXTASIS experiment \cite{antony}, where the observation frequency is
less than 5~MHz ($\lambda > 60$ m). When antennas are located at few hundreds of meters away
from the shower core, the exact field emitted by the shower particles hit the ground cannot be obtained
by adding the direct and reflected field since geometrical optics no longer apply.

What is the interest, then, of detecting cosmic rays at low frequency? 
Several experiments, such as Akeno \cite{r18} and EAS-TOP \cite{r23} have measured
an important low-frequency emission from cosmic ray showers (see \cite{benoiticrc} for a complete
review). At low frequencies, the
coherence of the shower is expected to be greater, giving rise to an electric field that has more
reach than in the standard [20-80] MHz band. This larger reach at low frequency has been
partially confirmed by EXTASIS \cite{antony}.
Another important feature at low frequency is the prediction of a pulse created by the coherent
deceleration of the shower particles when the shower particles hit the ground, called the
\emph{sudden death pulse} (SDP) \cite{sdp}. This pulse has not been seen by EXTASIS, although it might
be detected by a low-frequency experiment at a higher altitude. Besides, 
in \cite{sdp,sdparxiv,benoiticrc}, the SDP
has been approximated by the direct emission only, disregarding the effect of the interface.
We present in this work a complete approach that includes the effect of the boundary.

Another physical case for which the effect of the boundary on the electric field is important
corresponds to the detection of neutrinos in ice, with experiments like the aforementioned
ARA and ARIANNA. Whether the antennas are located over the ice or inside the ice, neutrino-induced
showers can cross the air/ice interface, and therefore the antennas will receive field coming
from both sides of the boundary. In the far field, the problem can be interpreted as a
transition radiation problem and the resulting electric field can be calculated using the far-field
formula for a track as a basis to obtain the direct, reflected and transmitted fields \cite{motloch,endpoints}.
Nevertheless, if the antenna and the particle track are close to the surface, the far-field approach
is not valid, in principle. It would be desirable, then, to have an exact calculation to employ when the
far-field approach breaks down, and also to assess the range of validity of the far-field approach.

In the present work, we calculate the field for a simple dipole and then we
increase the level of complexity to a particle track first and then to a shower toy model.
We begin deriving in Section~\ref{sec:dipole}
the electric field in frequency domain for an electric dipole embedded
in a space having two semi-infinite homogeneous media separated by a planar boundary.
In Section~\ref{sec:track} we use the field of a dipole to construct 
the exact electric field in frequency domain 
for a particle track inside the same two semi-infinite media. Then, we prove how 
it reduces to the formula in \cite{sdp} if there is no boundary, and as a consequence it can be
approximated by the ZHS formula \cite{zhs} in the far field. After that, we compare
the complete formula with a decomposition into direct, reflected, and transmitted fields,
both for air/soil and air/ice boundaries. We show as well that the field of a particle
inside soil can be neglected.
Finally, in Section~\ref{sec:model} 
we propose a simplified model for an air shower and show that the exact calculation
confirms (at least at a theoretical level) the existence of a pulse created by 
the deceleration of particles at ground
level (the SDP), which was already predicted in a less rigorous fashion by ignoring
the air/soil interface.

\section{Field of a dipole in two semi-infinite media separated by a planar boundary}
\label{sec:dipole}

Let us assume a three-dimensional space divided by a planar boundary at $z = 0$.
We define the upper region ($z > 0$) as medium~1, and the lower region ($z < 0$) as
medium~2. Both media are non-magnetic, that is, their permeability is equal to $\mu_0$,
the vacuum permeability.
Each medium has a relative permittivity $\epsilon_{jr}$ ($j = 1,2$), that can present
an imaginary part, indicating absorption. We assume that both media have
a conductivity $\sigma_j$. Using the $e^{-i \omega t}$ time dependence for the Fourier
transform \footnote{Or, equivalently, defining the Fourier transform as 
$\tilde{f}(\omega) = \int f(t) e^{i\omega t} \ \mathrm{d}t$.}, the
wavenumber is equal to
\begin{equation}
k_j = \omega \sqrt{\mu_0} \left[ \epsilon_0 \epsilon_{jr} + i \frac{\sigma_j}{\omega} \right]^{1/2},
\label{eq:wavenumber}
\end{equation}
where $\epsilon_0$ corresponds to the vacuum permittivity. We demand that the complex square root in
Eq.~\eqref{eq:wavenumber} lies in the upper part of the complex plane, so that the imaginary
part of $k_j$ is always positive or zero ($\Im(k_j) \geqslant 0$).

Let us now consider a vertical unit dipole vibrating at a given frequency $\omega$,
with a current density expressed in the following way:
\begin{equation}
\mathbf{J}\xw = \hat z \delta(x)\delta(y)\delta(z-z')\delta(\omega),
\label{eq:dipolecurrentdelta}
\end{equation}
where $z'$ is the vertical coordinate of the dipole.
Note that we have set the dipole moment equal to $1$~A\,m, which will be convenient later
for writing the field of a particle track. We can drop the $\delta(\omega)$ factor in
Eq.~\eqref{eq:dipolecurrentdelta} knowing that it must multiply the resulting dipole field
(since the dipole vibrates at one given frequency $\omega$ only), and write the new
dipole current as
\begin{equation}
\mathbf{J}\xw = \hat z \delta(x)\delta(y)\delta(z-z')
\label{eq:dipolecurrent}
\end{equation}

Although one can calculate the electromagnetic potentials or the Hertz potential and derivate
to obtain the electric and magnetic fields, for the general case of a dipole in an arbitrary direction
it is actually simpler to operate directly with Maxwell's equations expressed as a function of
the electric and magnetic fields. The main reason is that 
the enforcement of the boundary conditions can get a bit cumbersome using the potential
formalism, while they are naturally
expressed in terms of the fields. We begin by transforming Maxwell's equations to the
Fourier space $\kw,$ with the transform defined as:
\begin{equation}
\mathbf{A} \kw = \int_{-\infty}^{\infty} \ \mathrm{d}x \ \mathrm{d}y \ \mathrm{d} t
\ e^{i\omega t - ik_x x - i k_y y} \mathbf{A} \xt
\end{equation}
With this definition, we transform the horizontal coordinates $(x,y)$ to the Fourier space
$(k_x,k_y)$ and we switch from time domain to frequency domain as well. The vertical
coordinate $z$ remains untouched. Defining the $\tilde\nabla$ as
\begin{equation}
\tilde\nabla = ik_x \hat x + i k_y \hat y + \frac{\partial}{\partial z} \hat z,
\end{equation}
Maxwell's equations in $\kw$ space are formally identical to the standard equations in
frequency domain.
\begin{eqnarray}
\tilde\nabla \times \mathbf{E}_j & = & i \omega \mathbf{B}_j \nonumber \\
\tilde\nabla \times \mathbf{B}_j & = & -\frac{i k_j^2}{\omega}\mathbf{E}_j + \mu_0 \mathbf{J}
\label{eq:maxwell}
\end{eqnarray}
With $\mathbf{J}$ given by Eq.~\eqref{eq:dipolecurrent}. The subscript $j$ indicates the
upper (1) or lower (2) half-space. Eqs.~\eqref{eq:maxwell} can be decomposed into components
and combined in order to obtain solutions given in terms of imaginary exponentials. After that,
the resulting electric fields and magnetic fields are required to be 
continuous. In our case, magnetic fields must be continuous since both media are
non-magnetic. The tangential electric field must be continuous as well. The
normal component of the electric field must present a discontinuity. If we define the
complex electric displacement as the regular electric displacement for a dielectric medium
added to a term with the ohmic charge density:
\begin{equation}
\tilde{\mathbf{D}} = \epsilon\mathbf{E} + \frac{i}{\omega} \sigma \mathbf{E} =
\epsilon \mathbf{E} + \frac{i}{\omega}\mathbf{J}_\mathrm{ohm} \equiv \tilde\epsilon\mathbf{E},
\end{equation}
which implies that the divergence of $\tilde{\mathbf{D}}$ depends only on the free, non-ohmic current:
\begin{equation}
\nabla \cdot \tilde{\mathbf{D}} = \nabla(\epsilon\mathbf{E})
+ \frac{i}{\omega} \nabla\cdot\mathbf{J}_\mathrm{ohm} =
\rho_f - \rho_\mathrm{ohm} = \rho_{f,\mathrm{n-ohm}},
\label{eq:divD}
\end{equation}
where $\rho_{f,\mathrm{n-ohm}}$ is the free, non-ohmic charge density, $\rho_f$ 
is the charge density, and $\rho_\mathrm{ohm}$ is the ohmic current. We have used
the continuity equation for the ohmic current: 
\begin{equation}
-i\omega\rho_\mathrm{ohm} + \nabla\cdot\mathbf{J}_\mathrm{ohm} = 0.
\end{equation}
Since the only free, non-ohmic current or charge density present in the configuration is the 
one given by the dipole,
Eq.~\eqref{eq:divD} implies that the normal component of the electric field at the boundary,
whether in $\xw$ space or $\kw$ space must verify the condition:
\begin{equation}
\tilde\epsilon_1 E_{1z} - \tilde\epsilon_2 E_{2z}\Big|_{z=0} = k_1^2 E_{1z}- 
k_2^2 E_{2z}\Big|_{z=0} = 0.
\end{equation}

The calculation of the fields in each half-space and the enforcing of the boundary conditions
can be found in \cite{king}. We will adapt their formulas to our case. Throughout this
paper, we will assume that the observer lies in the upper half-space and will calculate the
electric field from a source located in the upper and the lower half-space. Of course, if the observer
lies in the lower half-space, the fields can be obtained by inverting the $z$ coordinate and swapping
the two media.

\subsection{Electric field from a vertical dipole located in the upper half-space}

\begin{figure}
\includegraphics[width=0.7\textwidth]{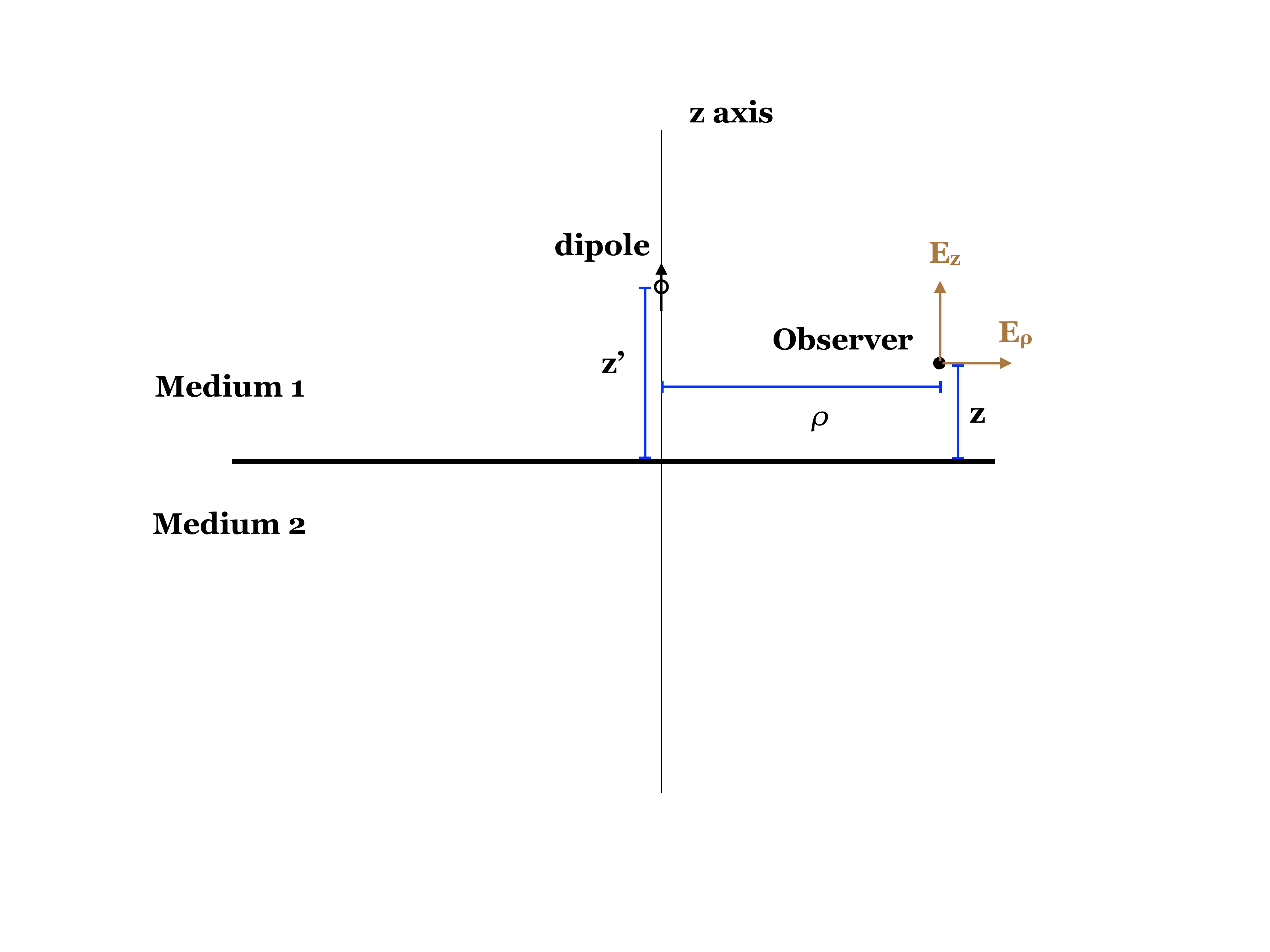}
\caption{Sketch of the geometry for the dipole electric field. The figure shows a plane with a
fixed azimuthal angle $\varphi$. The dipole is located on the $z$ axis, at $z'$. The observer lies
at a radial distance $\rho$ and a height $z$. The arrows indicate the radial and vertical components
of the electric field.}
\label{fig:sketch0}
\end{figure}

If both the observer and the dipole are located in medium 1 
($z > 0$, see Fig.~\eqref{fig:sketch0}), the electric
field can be written as a sum of direct field and the field created by the boundary.
The boundary field can be decomposed into the field created by a perfect image and
an integral involving Bessel functions:
\begin{equation}
\mathbf{E}_{v1}\xw = \mathbf{E}_\mathrm{direct} + \mathbf{E}_\mathrm{boundary} =
\mathbf{E}^d_{v1} + \mathbf{E}^{im}_{v1} + \mathbf{E}^{int}_{v1}
\label{eq:Ev}
\end{equation}
We assume the dipole is located at $(0,0,z' \geqslant 0)$ and the observer at
$(\rho\cos\varphi, \rho\sin\varphi, z)$. We define $r_1$ as the distance between
the dipole and the observer.
The radial and vertical components of the direct field can be written as \cite{king}:
\begin{eqnarray}
E^{d}_{1\rho} & = & -\frac{\omega\mu_0}{4\pi k_1^2} e^{i k_1 r_1}
\left( \frac{ik_1^2}{r_1} - \frac{3k_1}{r_1^2} - \frac{3i}{r_1^3}  \right)
\left( \frac{\rho}{r_1} \right) \left( \frac{z-z'}{r_1} \right) \nonumber \\
E^{d}_{1z} & =  &\frac{\omega \mu_0}{4\pi k_1^2} e^{i k_1 r_1}
\left[ \frac{ik_1^2}{r_1} - \frac{k_1}{r_1^2} - \frac{i}{r_1^3}
- \left( \frac{z-z'}{r_1} \right)^2 \left( \frac{ik_1^2}{r_1} - \frac{3k_1}{r_1^2}
- \frac{3i}{r_1^3} \right) \right]
\label{eq:Evd}
\end{eqnarray}
Defining $r_2 = \sqrt{\rho^2 + (z+z')^2}$, that is, the distance between the observer and the
dipole image at $(0,0,-z')$, the image field is readily written:
\begin{eqnarray}
E^{im}_{1\rho} & = & -\frac{\omega\mu_0}{4\pi k_1^2} e^{i k_1 r_2}
\left( \frac{ik_1^2}{r_2} - \frac{3k_1}{r_2^2} - \frac{3i}{r_2^3}  \right)
\left( \frac{\rho}{r_2} \right) \left( \frac{z+z'}{r_2} \right) \nonumber \\
E^{im}_{1z} & =  &\frac{\omega \mu_0}{4\pi k_1^2} e^{i k_1 r_2}
\left[ \frac{ik_1^2}{r_2} - \frac{k_1}{r_2^2} - \frac{i}{r_2^3}
- \left( \frac{z+z'}{r_2} \right)^2 \left( \frac{ik_1^2}{r_2} - \frac{3k_1}{r_2^2}
- \frac{3i}{r_2^3} \right) \right]
\label{eq:Evim}
\end{eqnarray}
The remaining part of the field is expressed with the help of an integral containing a Bessel function \cite{king}:
\begin{eqnarray}
E^{int}_{1\rho} & = & \frac{i\omega\mu_0 k_2^2}{2\pi k_1^2}
\int_0^\infty \frac{\gamma_1 e^{i\gamma_1(z+z')}}{N} J_1(k_\rho \rho)
k_\rho^2 \ \mathrm{d} k_\rho \nonumber \\
E^{int}_{1z} & = & -\frac{\omega\mu_0 k_2^2}{2\pi k_1^2}
\int_0^\infty \frac{ e^{i\gamma_1(z+z')}}{N} J_0(k_\rho \rho) k_\rho^3 \ \mathrm{d}k_\rho
\label{eq:Evint}
\end{eqnarray}
$k_\rho$ is an integration variable, $J_0$ and $J_1$ are the Bessel functions of the first kind
of zeroth and first order. $\gamma_j$, with $j = 1,2$ is defined as
\begin{equation}
\gamma_j \equiv \sqrt{k_j^2 - k_\rho^2},
\label{eq:gamma}
\end{equation}
and N is equal to
\begin{equation}
N \equiv k_1^2 \gamma_2 + k_2^2 \gamma_1
\label{eq:N}
\end{equation}
Eqs.~\eqref{eq:Evd}, \eqref{eq:Evim} and \eqref{eq:Evint} must be added to obtain the total
field, as in Eq.~\eqref{eq:Ev}. While the direct and image field are expressed in a closed form,
the integral in Eq.~\eqref{eq:Evint} is not expressable in a closed form in general, 
and must be computed numerically.
Several approximations can be made in order to simplify it. For instance, one can assume \cite{king}
that $|k_1| \gg |k_2|$, $\rho \geqslant 5z$, and $\rho \geqslant 5z'$, but the first of these conditions
is not verified when medium 1 is the atmosphere. We will show that another useful approximation
consists in using the direct field (or the ZHS formula) 
paired to the Fresnel coefficients when the observer lies
in the far field ($\Re(k_1) r_1 \gg 1$ and $\Re(k_1) r_2 \gg 1$). See Section~\ref{sec:fresnel}.

Due to the oscillatory nature and long tail of the Bessel functions and the imaginary exponential,
the integrals in Eq.~\eqref{eq:Evint} present a very slow convergence, and partition extrapolation
methods have to be used to keep the computation time reasonable. We explain it in 
Section~\ref{sec:partext}.

\subsection{Electric field from a vertical dipole located in the lower half-space}
\label{sec:Evunder}

Let us place the dipole now at $(0,0,z'<0)$, in the lower half-space, while retaining the observer
at $(\rho\cos\varphi, \rho\sin\varphi, z>0)$. The radial and vertical components of the electric field
$\mathbf{E}_{2\rightarrow 1}$ in such a case can be written as \cite{king}:
\begin{eqnarray}
E_{2\rightarrow 1,\rho} & = & - \frac{i\omega\mu_0}{2\pi}
\int_0^\infty \frac{\gamma_1 e^{-\gamma_2 z'} e^{i\gamma_1 z}}{N} J_1(k_\rho \rho)
k_\rho^2 \ \mathrm{d} k_\rho \nonumber \\
E_{2\rightarrow 1,z} & = & \frac{\omega\mu_0}{2\pi}
\int_0^\infty \frac{ e^{-i\gamma_2 z'} e^{i\gamma_1 z} }{N} J_0(k_\rho \rho) k_\rho^3 \ \mathrm{d}k_\rho.
\label{eq:Evunder}
\end{eqnarray}
Eq.~\eqref{eq:Evunder} corresponds to Eqs.~(3.3.13) and (3.3.14) in \cite{king} performing
a rotation of $\pi$ around the $x$ axis, which implies the
changes $E_z \rightarrow -E_z$, $z \rightarrow -z$, 
$z' \rightarrow -z'$,
and $\gamma_1 \rightarrow \gamma_2$. As opposed to Eq.~\eqref{eq:Ev},
Eq.~\eqref{eq:Evunder} implies that the field from a dipole in the lower half-space can be
written as a single integral with no evident closed form, although we will show in Section~\ref{sec:fresnel}
that in the far field
it is equivalent to the direct field (or ZHS formula) paired with the Fresnel coefficients.

\subsection{Electric field from a horizontal dipole located in the upper half-space}

Having obtained the field for a vertical dipole, we can solve Maxwell's equations
for a horizontal dipole. 
Instead of Eq.~\eqref{eq:dipolecurrent}, we choose the following current
along the x axis:
\begin{equation}
\mathbf{J}\xw = \hat x \delta(x)\delta(y)\delta(z-z'),
\label{eq:dipolecurrenth}
\end{equation}
which is the current of a horizontal unit dipole. The obtention of the field is analogous to
the vertical dipole case, and can be found in \cite{king}. Once again, the field for the
horizontal dipole in medium 1 when the observer is in the same medium 
can be expressed as a sum of three terms.
\begin{equation}
\mathbf{E}_{h1}\xw = \mathbf{E}_\mathrm{direct} + \mathbf{E}_\mathrm{boundary} =
\mathbf{E}^d_{h1} + \mathbf{E}^{im}_{h1} + \mathbf{E}^{int}_{h1}
\label{eq:Eh}
\end{equation}
In this case, however, for an observer at $(\rho\cos\varphi,\rho\sin\varphi,z)$, the azimuthal
component of the field is not zero, in general. The direct field is expressed as \cite{king}:
\begin{eqnarray}
E^{d}_{1\rho} & = & \frac{\omega\mu_0}{4\pi k_1^2} \cos\varphi
e^{ik_1 r_1} \left[ \frac{2k_1}{r_1^2} + \frac{2i}{r_1^3} + 
\frac{(z-z')^2}{r_1^2} \left( \frac{i k_1^2}{r_1} - \frac{3k_1}{r_1^2}
- \frac{3i}{r_1^3} \right) \right]
 \nonumber \\
 E^{d}_{1\varphi} & =  & - \frac{\omega\mu_0}{4\pi k_1^2} \sin\varphi
 e^{ik_1 r_1} \left( \frac{i k_1^2}{r_1} - \frac{k_1}{r_1^2}
- \frac{i}{r_1^3} \right)
 \nonumber \\
E^{d}_{1z} & =  &-\frac{i\omega \mu_0}{4\pi k_1^2} \cos\varphi e^{i k_1 r_1}
\left( \frac{\rho}{r_1} \right) \left( \frac{z-z'}{r_1} \right)
\left( \frac{k_1^2}{r_1} + \frac{3ik_1}{r_1^2}
- \frac{3}{r_1^3} \right).
\label{eq:Ehd}
\end{eqnarray}
Eq.~\eqref{eq:Ehd} can be obtained from the direct field for the vertical dipole
(Eq.~\eqref{eq:Evd}) after perfoming two rotations, which shows that the calculations
are consistent (see Appendix~\ref{sec:vtoh}).
The image field can be written as
\begin{eqnarray}
E^{im}_{1\rho} & = & \frac{\omega\mu_0}{4\pi k_1^2} \cos\varphi
e^{ik_1 r_2} \left[ \frac{2k_1}{r_2^2} + \frac{2i}{r_2^3} + 
\frac{(z+z')^2}{r_2^2} \left( \frac{i k_1^2}{r_2} - \frac{3k_1}{r_2^2}
- \frac{3i}{r_2^3} \right) \right]
 \nonumber \\
 E^{im}_{1\varphi} & =  & - \frac{\omega\mu_0}{4\pi k_1^2} \sin\varphi
 e^{ik_1 r_2} \left( \frac{i k_1^2}{r_2} - \frac{k_1}{r_2^2}
- \frac{i}{r_2^3} \right)
 \nonumber \\
E^{im}_{1z} & =  &-\frac{i\omega \mu_0}{4\pi k_1^2} \cos\varphi e^{i k_1 r_2}
\left( \frac{\rho}{r_2} \right) \left( \frac{z+z'}{r_2} \right)
\left( \frac{k_1^2}{r_2} + \frac{3ik_1}{r_2^2}
- \frac{3}{r_2^3} \right).
\label{eq:Ehim}
\end{eqnarray}
The integrals for the horizontal case are slightly more complicated than the ones
found for the vertical case \cite{king}:
\begin{eqnarray}
E^{int}_{1\rho} = - \frac{\omega\mu_0}{4\pi k_1^2} \cos\varphi
\int_0^\infty & \Bigg( & \frac{\gamma_1}{2} (Q-1) [J_0(k_\rho \rho)-J_2(k_\rho \rho)] \nonumber \\
& - & \frac{k_1^2}{2\gamma_1} (P+1) [J_0(k_\rho \rho)+J_2(k_\rho \rho)]
\Bigg) e^{i\gamma_1(z+z')} k_\rho \ \mathrm{d} k_\rho \nonumber \\
E^{int}_{1\varphi} = \frac{\omega\mu_0}{4\pi k_1^2} \sin\varphi
\int_0^\infty & \Bigg( & \frac{\gamma_1}{2} (Q-1) [J_0(k_\rho \rho)+J_2(k_\rho \rho)] \nonumber \\
& - & \frac{k_1^2}{2\gamma_1} (P+1) [J_0(k_\rho \rho)-J_2(k_\rho \rho)]
\Bigg) e^{i\gamma_1(z+z')} k_\rho \ \mathrm{d} k_\rho \nonumber \\
E^{int}_{1z} = \frac{i \omega\mu_0}{2\pi k_1^2} \cos\varphi
\int_0^\infty & & (Q-1) J_1(k_\rho \rho) e^{i\gamma_1(z+z')} k_\rho^2
 \ \mathrm{d}k_\rho,
\label{eq:Ehint}
\end{eqnarray}
with
\begin{equation}
P \equiv \frac{\gamma_2-\gamma_1}{\gamma_2+\gamma_1},
\end{equation}
and
\begin{equation}
Q \equiv \frac{k_1^2\gamma_2 - k_2^2\gamma_1}{k_1^2\gamma_2 + k_2^2\gamma_1}
\end{equation}

\subsection{Electric field from a horizontal dipole located in the lower half-space}

With the same geometry as in Section~\ref{sec:Evunder} the field from a horizontal
dipole in medium 2 seen by an observer in medium 1 is:
\begin{eqnarray}
E_{2\rightarrow 1,\rho} = - \frac{\omega\mu_0}{4\pi} \cos\varphi
\int_0^\infty & \Bigg( &  M^{-1} [J_0(k_\rho \rho) + J_2(k_\rho \rho)] \nonumber \\ 
& + & \frac{\gamma_1 \gamma_2}{N} [J_0(k_\rho \rho) - J_2(k_\rho \rho)] \Bigg)
e^{i(-\gamma_2 z' + \gamma_1 z)}
k_\rho^2 \ \mathrm{d} k_\rho \nonumber \\
E_{2\rightarrow 1,\varphi} = \frac{\omega\mu_0}{4\pi} \sin\varphi
\int_0^\infty & \Bigg( & M^{-1} [J_0(k_\rho \rho) - J_2(k_\rho \rho)] \nonumber \\
& + & \frac{\gamma_1 \gamma_2}{N} [J_0(k_\rho \rho) + J_2(k_\rho \rho)] \Bigg)
e^{i(-\gamma_2 z' + \gamma_1 z)}
k_\rho \ \mathrm{d} k_\rho \nonumber \\
E_{2\rightarrow 1,z} = \frac{i\omega\mu_0}{2\pi} \cos\varphi
\int_0^\infty & & \frac{\gamma_2}{N} J_1(k_\rho \rho)
e^{i(-\gamma_2 z' + \gamma_1 z)}
k_\rho^2 \ \mathrm{d} k_\rho 
\label{eq:Ehunder}
\end{eqnarray}
Eq.~\eqref{eq:Evunder} corresponds to Eqs.~(5.4.35), (5.4.36) and (5.4.37) in \cite{king} performing
a rotation of $\pi$ around the $x$ axis, which implies the
changes $E_z \rightarrow -E_z$, $z \rightarrow -z$,  
$z' \rightarrow -z'$, $E_\varphi \rightarrow -E_\varphi$, $\varphi \rightarrow -\varphi$,
and $\gamma_1 \rightarrow \gamma_2$. $M$ is defined as:
\begin{equation}
M = \gamma_1 + \gamma_2
\end{equation}

\subsection{Evaluation of Bessel integrals}
\label{sec:partext}

Due to the presence of ordinary Bessel functions $J_n$, the integrals in Eqs.~\eqref{eq:Evint},
\eqref{eq:Evunder}, \eqref{eq:Ehint} and \eqref{eq:Ehunder} contain long oscillating tails that
are difficult to evaluate numerically while keeping the computation time reasonable. A slow but
easy way to calculate them is to compute the subintegrals in several intervals of the real number line:
[$0,a_1$], [$a_1, a_2$], [$a_2, a_3$], etc, and stop when the sum of the subintegrals reaches
convergence with the desired precision. This method is computationally intensive, 
and the evaluation time of a single integral is of the order of a few seconds. 

In order to compute the electric fields in a faster way, we have used an extrapolation method
called the partition extrapolation method \cite{partextr}. Our application of this method goes
as follows. We take the intervals $[0,b]$, $[b,2b]$, $[2b, 3b]$, ..., $[(n-1)b, nb]$, and
we calculate the integrals of the function in each interval, $I_1$, $I_2$, ..., $I_n$,
using a quadrature adaptative
method such as QAG from the GNU Scientific Library (also available via Scipy in Python).
Once the value of these integrals is known, the partial sums for each interval are needed, defined as:
\begin{equation}
S_k = \sum_{i = 1}^{k} I_i.
\end{equation}
Let us order these partial sums, $[S_1, S_2, ..., S_n]$. The partition extrapolation method tells us that we
can obtain an estimation of the [$0,\infty$) integral operating on this array by using a triangular scheme.
Starting from the array $[S_1, S_2, ..., S_n]$
containing the partial sums, we define another array with $n-1$ elements, and where
each element is an average of two contiguous elements of our original array.
\begin{equation}
\left[ \frac{1}{2}(S_1+S_2), \frac{1}{2}(S_2+S_3), ..., \frac{1}{2}(S_{n-1}+S_{n}) \right]
\equiv [S^{(1)}_1, S^{(1)}_2, ..., S^{(1)}_{n-1}]
\end{equation}
Each iteration can be computed, then, knowing that for the step $k$, the element $i$ is calculated
using the elements from the previous step $k-1$,
\begin{equation}
S^{(k)}_i = \frac{1}{2} (S^{(k-1)}_i + S^{(k-1)}_{i+1})
\end{equation}
After $n-1$ iterations we end up with an array containing a single element, $S^{(n-1)}_1$. This element
is an estimation for the Bessel integral, and this estimation converges much faster than the
brute-force approach \cite{partextr}. 
The main idea behind the partition extrapolation method is that the averaging of the partial sums tends
to dampen the oscillations of the Bessel function, while converging towards the true value of the
integral. The partition extrapolation method allows the use, in general,
of different weights for the averaging of the partial sums. However, we have chosen the simple $\frac{1}{2}$
factor because the different weights proposed in \cite{partextr} resulted in numerical unstabilities
for our particular integrals.

Depending on the wavenumber $k$ and the radial distance $\rho$, as well as on the desired precision,
a different number of intervals are needed. We will use $b = \pi/\rho$
for the period, while for the number of intervals, we have chosen:
\begin{equation}
n = \max\left( 25, 
\Big\lfloor 16 \frac{\nu}{1\ \mathrm{MHz}} \frac{\rho}{1\ \mathrm{km}} \Big\rfloor \right).
\end{equation}
$\nu$ represents the linear frequency. 
For certain observers too close
to the dipole, we will use the double of intervals and a period of $b/\sqrt{2}$.
With this convention, the relative error of the integral
is always less than $10^{-5}$ for the range of frequencies and distances explored throughout
this paper.

Once the solution for the field of a dipole is known, we can use it 
to obtain the complete field created by a particle track.

\section{Field of a particle track in two semi-infinite media separated by a planar boundary}
\label{sec:track}

Let us consider a particle with charge $q$ at rest at the point $\mathbf{x} = \mathbf{x}_1$. 
At a time $t = t_1$ 
the particle is suddenly accelerated and begins to travel in a straight line with a velocity $\mathbf{v}$. 
Then, at $t = t_2$ the particle is abruptly stopped and stays at that place. 
This trajectory is called a particle track, and it constitutes the building block for
particle physics Monte Carlo codes in general, and in particular for the codes that calculate the
radio emission from particle showers, such as SELFAS \cite{selfas}, ZHAireS \cite{zhaires},
CoREAS \cite{coreas}, or ZHS \cite{zhs}.
The current density of a particle track can be written in time domain as:
\begin{equation}
\mathbf{J}_\mathrm{track}\xt = q \mathbf{v} \deltav [\rect],
\label{eq:current}
\end{equation}
where $\Theta$ is the Heaviside step function.
We do not need the charge density for our calculations, since our frequency domain formulas
are complete using the current density only, \emph{i.e.} charge conservation is automatically
taken into account if we solve the curl equations in frequency (Eq.~\eqref{eq:maxwell}), as it
has been done for obtaining the dipole fields. Eq.~\eqref{eq:current} can be transformed to
frequency domain to yield:
\begin{equation}
\mathbf{J}_\mathrm{track}\xw = q\mathbf{v}  \int_{t_1}^{t_2} \mathrm{d}t' \ e^{i\omega t'}
\delta^3(\mathbf{x} - \mathbf{x}_1 -\mathbf{v}(t'-t_1) )
\label{eq:freqcurrent}
\end{equation}
The electric field created by this current can be calculated as a superposition of the
fields for a vertical and a horizontal unit dipole. Since a frame where the track velocity has
no $y$ component can always be found, we can assume without loss of generality:
\begin{equation}
\mathbf{v} = v_x \hat x + v_z \hat z \equiv v ( \cos\theta \hat x + \sin\theta \hat z )
\end{equation}
Eqs.~\eqref{eq:Evd}, \eqref{eq:Evim}, \eqref{eq:Evint}, \eqref{eq:Evunder},
\eqref{eq:Ehd}, \eqref{eq:Ehim}, \eqref{eq:Ehint}, \eqref{eq:Ehunder} for the fields have
been calculated for a dipole located at $(0,0,z')$, but they are still valid for a dipole
at $\mathbf{x'} = (x',0,z')$ provided the cylindrical coordinate system $(\rho,\varphi,z)$ is correctly
centered on the vertical axis that passes through the dipole position. Let
$\mathbf{E}_{v(h)}(\mathbf{x},\mathbf{x'}, \omega)$ 
be the field created by a vertical (horizontal) unit dipole at $\mathbf{x'} = (x',0,z')$. The field
for a particle track can be expressed as a combination of these fields:
\begin{equation}
\mathbf{E}_\mathrm{track} \xw = 
\frac{q v}{1~\mathrm{A}\cdot\mathrm{m}}\int_{t_1}^{t_2} \mathrm{d}t' \ e^{i\omega t'}
\left[
\cos\theta \mathbf{E}_{h}(\mathbf{x},\mathbf{x'}(t'), \omega) +
\sin\theta \mathbf{E}_{v}(\mathbf{x},\mathbf{x'}(t'), \omega)
\right],
\label{eq:etrack}
\end{equation}
where $\mathbf{x'}(t')$ is the trajectory of a particle track:
\begin{equation}
\mathbf{x'}(t') = \mathbf{x}_1 + (t'-t_1)(v_x \hat x + v_z \hat z)
\label{eq:xtprime}
\end{equation}
We have included a dimensional factor of $1\ \mathrm{A}\cdot\mathrm{m}$, since the unit
dipole possesses a moment of the same magnitude. We can define a magnetic field for a particle
track in the same way, making the substitution $E\rightarrow B$ in Eq.~\eqref{eq:etrack} and
taking $\mathbf{B}_{v(h)}$ as the magnetic field from a unit dipole:
\begin{equation}
\mathbf{B}_\mathrm{track} \xw = 
\frac{q v}{1~\mathrm{A}\cdot\mathrm{m}}\int_{t_1}^{t_2} \mathrm{d}t' \ e^{i\omega t'}
\left[
\cos\theta \mathbf{B}_{h}(\mathbf{x},\mathbf{x'}(t'), \omega) +
\sin\theta \mathbf{B}_{v}(\mathbf{x},\mathbf{x'}(t'), \omega)
\right]
\label{eq:btrack}
\end{equation}
$\mathbf{E}_{v(h)}(\mathbf{x},\mathbf{x'}, \omega)$ 
verifies the following Maxwell's equations:
\begin{eqnarray}
\nabla \times \mathbf{E}_{v,h}(\mathbf{x},\mathbf{x'}, \omega) & = & 
i \omega \mathbf{B}_{v,h}(\mathbf{x},\mathbf{x'}, \omega) \nonumber \\
\nabla \times \mathbf{B}_{v,h}(\mathbf{x},\mathbf{x'}, \omega) & = &
- \frac{i k_j^2}{\omega} \mathbf{E}_{v,h}(\mathbf{x},\mathbf{x'}, \omega)
+ \mu_0 \mathbf{J}_{v,h}(\mathbf{x},\mathbf{x}',\omega),
\label{eq:maxdip}
\end{eqnarray}
$j = 1,2$ and
where the currents are given by
\begin{eqnarray}
\mathbf{J}_{v}(\mathbf{x},\mathbf{x}',\omega) = \hat z \delta(x-x')\delta(y)\delta(z-z') \nonumber \\
\mathbf{J}_{h}(\mathbf{x},\mathbf{x}',\omega) = \hat x \delta(x-x')\delta(y)\delta(z-z'),
\end{eqnarray}
which means that the track current can be expressed by:
\begin{equation}
\mathbf{J}_\mathrm{track}\xw = q
\int_{t_1}^{t_2} \mathrm{d}t' \ e^{i\omega t'}
\left(
v_x \mathbf{J}_{h}(\mathbf{x},\mathbf{x}'(t'),\omega) + 
v_z \mathbf{J}_{v}(\mathbf{x},\mathbf{x}'(t'),\omega)
\right),
\label{eq:currcomb}
\end{equation}
with $\mathbf{x}'(t')$ taken from Eq.~\eqref{eq:xtprime}.
Eq.~\eqref{eq:maxdip} implies that the field of a track as defined in Eq.~\eqref{eq:etrack}
verifies Maxwell's equations as well and yields the correct current density (Eq.~\eqref{eq:freqcurrent}).
The equation for the curl of the electric field is trivially verified if we define the magnetic field
$\mathbf{B}_\mathrm{track} \xw$ as in Eq.~\eqref{eq:btrack}, since the dipole fields
verify $\nabla\times\mathbf{E}_{v(h)} = i\omega\mathbf{B}_{v(h)}$, and our track field is
a linear combination of the dipole fields.
As for the curl of the magnetic field,
\begin{eqnarray}
\nabla \times \mathbf{B}_\mathrm{track} \xw  & = &
\frac{q}{1\ \mathrm{A}\cdot\mathrm{m}}\int_{t_1}^{t_2} \mathrm{d}t' \ e^{i\omega t'}
\nabla \times \left[
v_x \mathbf{B}_{h}(\mathbf{x},\mathbf{x'}(t'), \omega) +
v_z \mathbf{B}_{v}(\mathbf{x},\mathbf{x'}(t'), \omega)
\right] \nonumber \\
& = &
\frac{q}{1\ \mathrm{A}\cdot\mathrm{m}}\int_{t_1}^{t_2} \mathrm{d}t' \ e^{i\omega t'}
\left[
-\frac{i k_j^2}{\omega} \left( v_x \mathbf{E}_{h} +
v_z \mathbf{E}_{v}\right) +
\mu_0 \left( v_x \mathbf{J}_{h} + 
v_z \mathbf{J}_{v} \right)
\right] \nonumber \\
& = & -\frac{i k_j^2}{\omega} \mathbf{E}_\mathrm{track} + 
\mu_0 \mathbf{J}_\mathrm{track},
\label{eq:checkmax}
\end{eqnarray}
where we have used Eq.~\eqref{eq:currcomb}. Eq.~\eqref{eq:checkmax} implies that our solution
for the electric field of a particle track (Eq.~\eqref{eq:etrack}) verifies Maxwell's equations, and
the source current is precisely the current of a particle track, as intended.

\subsection{Comparison with previous analytical calculations}

With Eq.~\eqref{eq:Evd} and \eqref{eq:etrack}, the radial component of the direct 
electric field created by a vertical track is:
\begin{equation}
E^{d}_{\mathrm{track},1\rho}  =  -\frac{q v\omega\mu_0}{4\pi k_1^2} 
\int_{t_1}^{t_1} \mathrm{d}t' \ e^{i\omega t'} e^{i k_1 r_1}
\left( \frac{ik_1^2}{r_1} - \frac{3k_1}{r_1^2} - \frac{3i}{r_1^3}  \right)
\left( \frac{\rho}{r_1} \right) \left( \frac{z-z'}{r_1} \right).
\label{eq:oldrho}
\end{equation}
$r_1 = r_1(t')$, the distance from particle to observer, is now a function of time.
Also, $z'(t) = z_1 + vt'$.
Analogously, for the $z$ component one finds:
\begin{equation}
E^{d}_{\mathrm{track},1z}  =  \frac{qv\omega \mu_0}{4\pi k_1^2}
\int_{t_1}^{t_1} \mathrm{d}t' \ e^{i\omega t'} e^{i k_1 r_1}
\left[ \frac{ik_1^2}{r_1} - \frac{k_1}{r_1^2} - \frac{i}{r_1^3}
- \left( \frac{z-z'}{r_1} \right)^2 \left( \frac{ik_1^2}{r_1} - \frac{3k_1}{r_1^2}
- \frac{3i}{r_1^3} \right) \right]
\label{eq:oldz}
\end{equation}
Eqs.~\eqref{eq:oldrho} and \eqref{eq:oldz} are the same, but with a different
notation, as Eqs.~(12) and (13) for a vertical
track found in \cite{prd87}. If we make $k_1^2 = \omega^2 \mu_0 \epsilon$, 
and $b = ik_1 - 1/r_1$, we arrive at the same expression, and
as a consequence the direct fields of the present work are completely equivalent.
This implies as well that they are equivalent to the fields in \cite{sdparxiv} and that
they yield the ZHS formula \cite{zhs} as a far-field approximation.

\subsection{Evaluation of the field for a single particle track}

After integrand in Eq.~\eqref{eq:etrack} is known, the integral from $t_1$ to $t_2$ must
be computed. Eq.~\eqref{eq:etrack} in conjunction with Eqs.~\eqref{eq:Ev} and \eqref{eq:Eh}
tells us that if observer and track are in medium 1, the electric field can be written as a
superposition of the direct field, the image field, and the integral field from the unit dipole:
\begin{equation}
\mathbf{E}_\mathrm{track} \xw = 
\frac{q v}{1~\mathrm{A}\cdot\mathrm{m}}\int_{t_1}^{t_2} \mathrm{d}t' \ e^{i\omega t'}
\left[
\mathbf{E}^{d}_{1} + \mathbf{E}^{im}_{1} + \mathbf{E}^{int}_{1}
\right]
\equiv \mathbf{E}_\mathrm{track}^{d} + \mathbf{E}_\mathrm{track}^{im} +
\mathbf{E}_\mathrm{track}^{int},
\label{eq:etracksum}
\end{equation}
having defined $E^{d,im,int}_1$ as a combination of the vertical and horizontal unit dipole fields,
\begin{equation}
\mathbf{E}^{d,im,int}_1 \equiv \cos\theta\mathbf{E}^{d,im,int}_{h,1} + 
\sin\theta\mathbf{E}^{d,im,int}_{v,1}.
\end{equation}
$\mathbf{E}^{d}_{v(h),1}$ is found in Eq.~\eqref{eq:Evd} (Eq.~\eqref{eq:Ehd}),
$\mathbf{E}^{im}_{v(h),1}$ in Eq.~\eqref{eq:Evim} (Eq.~\eqref{eq:Ehim}), and
$\mathbf{E}^{int}_{v(h),1}$ in Eq.~\eqref{eq:Evint} (Eq.~\eqref{eq:Ehint}).
The integral in Eq.~\eqref{eq:etracksum} can be numerically evaluated with a Riemann sum.
However, $\mathbf{E}^{d}_{v(h),1}$ and $\mathbf{E}^{im}_{v(h),1}$  exhibit a
$e^{i k r}$ dependence, which allows us to invoke the Fraunhofer approximation.
If the parameter 
\begin{equation}
\eta \equiv \frac{k_1 L^2}{r} \sin\theta_\mathrm{obs} \ll 1,
\label{eq:eta}
\end{equation}
where $L$ is the length of the track and $\sin\theta_\mathrm{obs}$ the angle between the
track velocity and the line joining the track and the observer, we can make the following
assumptions:
\begin{enumerate}
\item Fraunhofer approximation for the phase:
\begin{equation}
k r = k |\mathbf{x}-\mathbf{x'}| \approx k[R - v(t-t_0)\cos\theta_\mathrm{obs}],
\end{equation}
where R is the distance from the observation point to a reference point along the track where
the particle is located at a reference time $t_0$. For our analysis, the reference point taken
will be the middle point of the track.
\item We assume the distances from track to observer can be taken as constant (except
for the one present in the phase), or equivalently
\begin{equation}
\frac{L}{R} \ll 1.
\end{equation}
\end{enumerate}
If the above conditions are not fulfilled, we can always subdivide the track until they are.
With the above approximations, the integrals can be analytically computed, with the
result:
\begin{equation}
\mathbf{E}_\mathrm{track}^{d} \approx
\frac{q v}{1~\mathrm{A}\cdot\mathrm{m}} e^{i \mathbf{k}_d\cdot\mathbf{v} t_0}
\left[
\frac{e^{i (\omega - \mathbf{k}_d\cdot\mathbf{v}) t_2}-
e^{i (\omega - \mathbf{k}_d\cdot\mathbf{v}) t_1 }}{i(\omega - \mathbf{k}_d\cdot\mathbf{v})}
\right] \mathbf{E}^{d}_1(\mathbf{x},\mathbf{x'}_0, t_0),
\label{eq:edfraun}
\end{equation}
with the track position fixed at:
\begin{equation}
\mathbf{x'}_0 = \mathbf{x}_1 + \mathbf{v}(t_0-t_1),
\end{equation}
$t_0$ being:
\begin{equation}
t_0 = \frac{t_1+t_2}{2},
\end{equation}
and $\mathbf{k}_d$ defined as the wavenumber times the unit vector from $\mathbf{x'}_0$
to the observer in $\mathbf{x}$:
\begin{equation}
\mathbf{k}_d = k_1 \frac{\mathbf{x}-\mathbf{x'}_0}{|\mathbf{x}-\mathbf{x'}_0|}
\label{eq:wavevector}
\end{equation}
Eq.~\eqref{eq:edfraun} reduces to the ZHS formula when dropping all the terms that fall
faster than $\frac{1}{r}$ (far-field approximation). 
The Fraunhofer approximation can be used for the image field as
well, but keeping in mind that the expansion has to be made around the image point:
\begin{equation}
\mathbf{x'}_{0,im} = x_0 \hat x - z_0 \hat z,
\end{equation}
which is the mirror point of the middle of the track. This modifies as well the definition of
the wavenumber vector:
\begin{equation}
\mathbf{k}_{im} = k_1 \frac{\mathbf{x}-\mathbf{x'}_{0,im}}{|\mathbf{x}-\mathbf{x'}_{0,im}|}.
\end{equation}
Other than that, the Fraunhofer approximation for the image field is analogous to Eq.~\eqref{eq:edfraun}:
\begin{equation}
\mathbf{E}_\mathrm{track}^{im} \approx
\frac{q v}{1~\mathrm{A}\cdot\mathrm{m}} e^{i \mathbf{k}_{im}\cdot\mathbf{v} t_0}
\left[
\frac{e^{i (\omega - \mathbf{k}_{im}\cdot\mathbf{v}) t_2}-
e^{i (\omega - \mathbf{k}_{im}\cdot\mathbf{v}) t_1 }}{i(\omega - \mathbf{k}_{im}\cdot\mathbf{v})}
\right] \mathbf{E}^{im}_1(\mathbf{x},\mathbf{x'}_{0,im}, t_0),
\label{eq:eimfraun}
\end{equation}
Eqs.~\eqref{eq:edfraun} and \eqref{eq:eimfraun} can be as accurate numerical integration,
provided the Fraunhofer conditions are verified.

\begin{figure}
\includegraphics[width=0.8\textwidth]{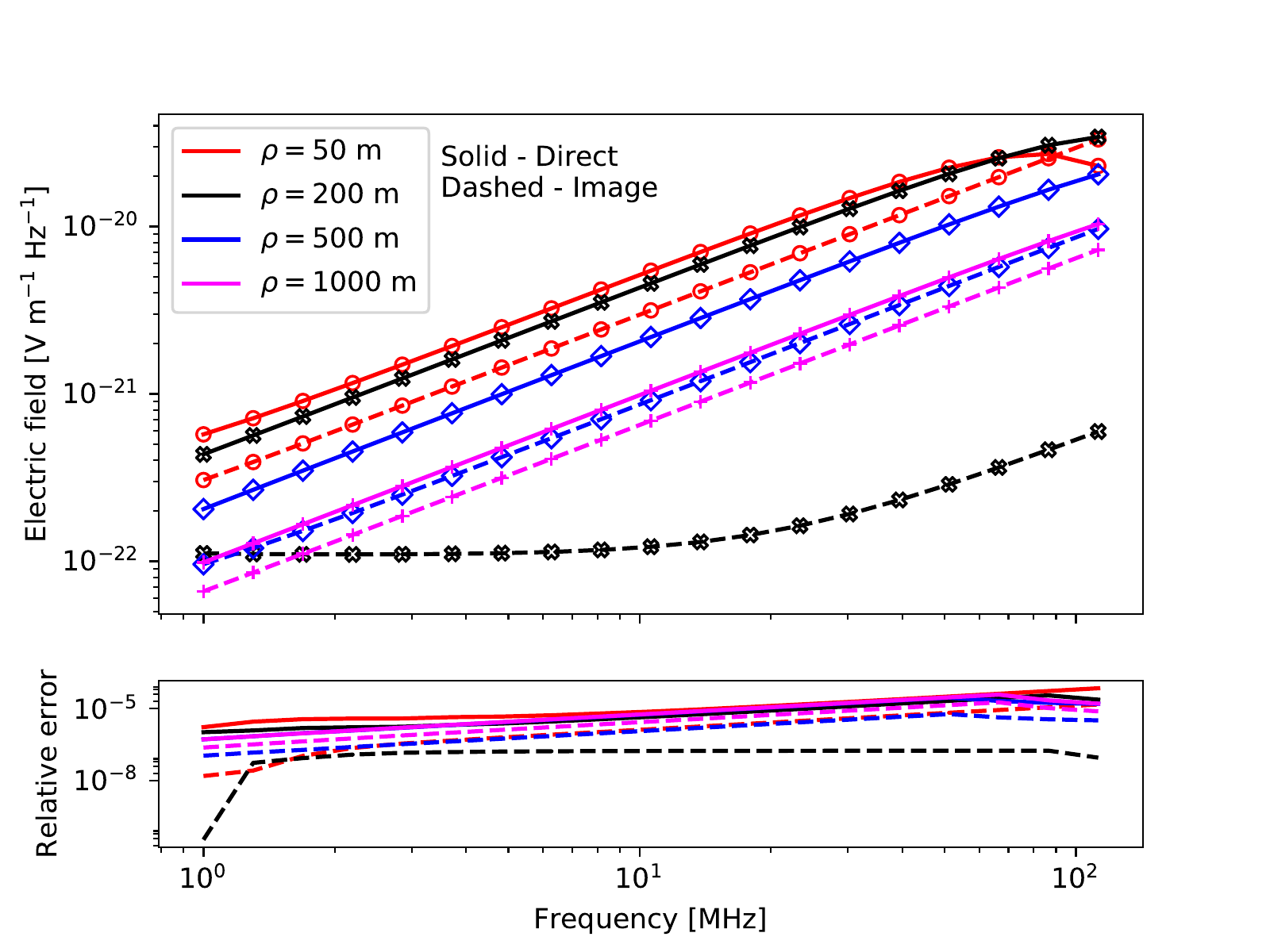}
\caption{Top: Electric field as a function of frequency created by a $1.2$~m long electron track
traveling at a speed $\sim c$. The middle point of the track is $5$~m above the interface
and the track is traveling towards the boundary with an angle of $\pi/4$ with respect to
the normal. The observers are located at $2$~m from the ground at radial distances of
50 (red circles), 200 (black crosses) 500~m (blue diamonds), and
1000~m (magenta plus symbols). Solid lines represent direct fields,
dashed lines represent image fields, and dash-dotted lines represent the integral fields.
Lines have been calculated using the Fraunhofer approximation and points have been
calculated performing a Riemann sum. 
Bottom: Relative error for the Fraunhofer approximation compared to the Riemann sum.
See text for details.}
\label{fig:frauncheck}
\end{figure}

Calculating $\mathbf{E}_\mathrm{track}^{int}$, in principle, needs to be done numerically.
We can use, for instance, a simple Riemann sum or a Gauss-Legendre quadrature method
for an appropriate number of subdivisions of the $[t_1,t_2]$ interval.
\begin{comment}
However, since Green's function for the Helmholtz equation for the electromagnetic potential 
contains a $e^{ikr}$ term homogeneous media regardless of the presence of a boundary,
we can make the hypothesis that the integral fields in Eqs.~\eqref{eq:Evint} and \eqref{eq:Ehint}
possess a similar dependence and therefore we can approximate the integral in the same
way as with the image field (Eq.~\eqref{eq:eimfraun}), with the same $\mathbf{x'}_{0,im}$
and $\mathbf{k}_{im}$:
\begin{equation}
\mathbf{E}_\mathrm{track}^{int} \approx
\frac{q v}{1~\mathrm{A}\cdot\mathrm{m}} e^{i \mathbf{k}_{im}\cdot\mathbf{v} t_0}
\left[
\frac{e^{i (\omega - \mathbf{k}_{im}\cdot\mathbf{v}) t_2}-
e^{i (\omega - \mathbf{k}_{im}\cdot\mathbf{v}) t_1 }}{i(\omega - \mathbf{k}_{im}\cdot\mathbf{v})}
\right] \mathbf{E}^{int}_1(\mathbf{x},\mathbf{x'}_{0,im}, t_0),
\label{eq:eintfraun}
\end{equation}
\end{comment}
We show in Fig.~\ref{fig:frauncheck} that the Fraunhofer approximation for the integral is
appropriate for the direct and image fields. We have placed a $1.2$~m long electron track whose
medium point is at $5$ m from the interface, heading for the ground with a $\pi/4$ angle
with respect to the vector normal to the interface. 
Medium 1 is air with $\epsilon_r = (1.0001)^2$ and no conductivity, and medium 2 is
an average soil at the Nan\c{c}ay Observatory, where the EXTASIS experiment is located,
having $\epsilon_r = 12$ and a conductivity of $\sigma = 5$~mS/m.
We have placed several observers $2$ m above the boundary and at radial distances
of 50, 200, and 500~m.
The direct and image fields have been calculated using a Riemann sum after dividing
the integrand in intervals of $T/1000$, where $T$ is the period for each frequency, and also
with the Fraunhofer approximation by enforcing $\eta < 10^{-2}$ (Eq.~\eqref{eq:eta}).
On top of Fig.~\ref{fig:frauncheck} we have plotted the modules of the electric field for the
direct and image fields (points represent the Riemann sum and lines the Fraunhofer
approximation), and on the bottom the relative error the Fraunhofer approximation.
The agreement is better than $< 10^{-4}$, which is more than satisfactory for our purposes.
We have checked that the complex components of the fields agree, and not only the modules,
which is vital to correctly account for interference between particle fields. 
All of this means that we can use either Eqs.~\eqref{eq:edfraun} and
\eqref{eq:eimfraun}, subdividing when necessary or 
integrating with whatever numerical method we prefer.
The approximation is still valid, as it should, when the track or the observer are far from the boundary. 

If the track is in medium 2 and the observer in medium 1, the transmitted field must be obtained
with the help of Eqs.~\eqref{eq:Evunder} and \eqref{eq:Ehunder}:
\begin{equation}
\mathbf{E}_\mathrm{track,2\rightarrow 1} \xw = 
\frac{q v}{1~\mathrm{A}\cdot\mathrm{m}}\int_{t_1}^{t_2} \mathrm{d}t' \ e^{i\omega t'}
\left[
\cos\theta \mathbf{E}_{h, 2\rightarrow 1} + \sin\theta \mathbf{E}_{v, 2\rightarrow 1}
\right].
\label{eq:etracksumunder}
\end{equation}
We will integrate Eq.~\eqref{eq:etracksumunder} numerically, as with $\mathbf{E}_\mathrm{track}^{int}$.
%We have found that, for Eq.~\eqref{eq:etracksumunder}, the Fraunhofer approximation is
%not that good for frequencies higher than a few tens of MHz. The reason is probably related
%to the refraction, which changes the direction and the distance covered by a wave
%in a different way than for a direct or image field, and therefore we cannot suppose
%$e^{ikr}$ dependence for the expansion. As a consequence, we will integrate 

\subsection{Direct, boundary, and total fields}

\begin{figure}
\includegraphics[width=0.7\textwidth]{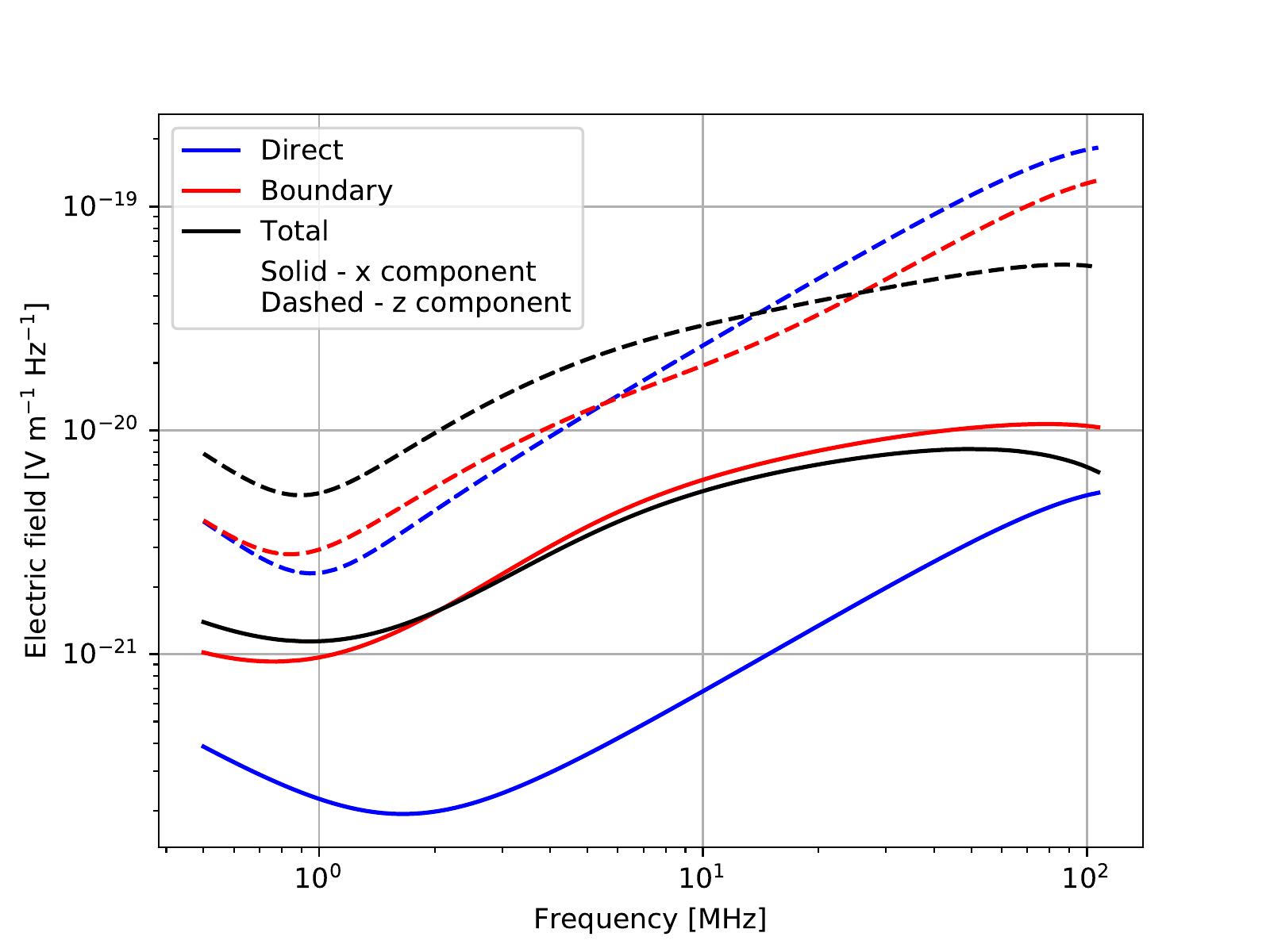}
\includegraphics[width=0.7\textwidth]{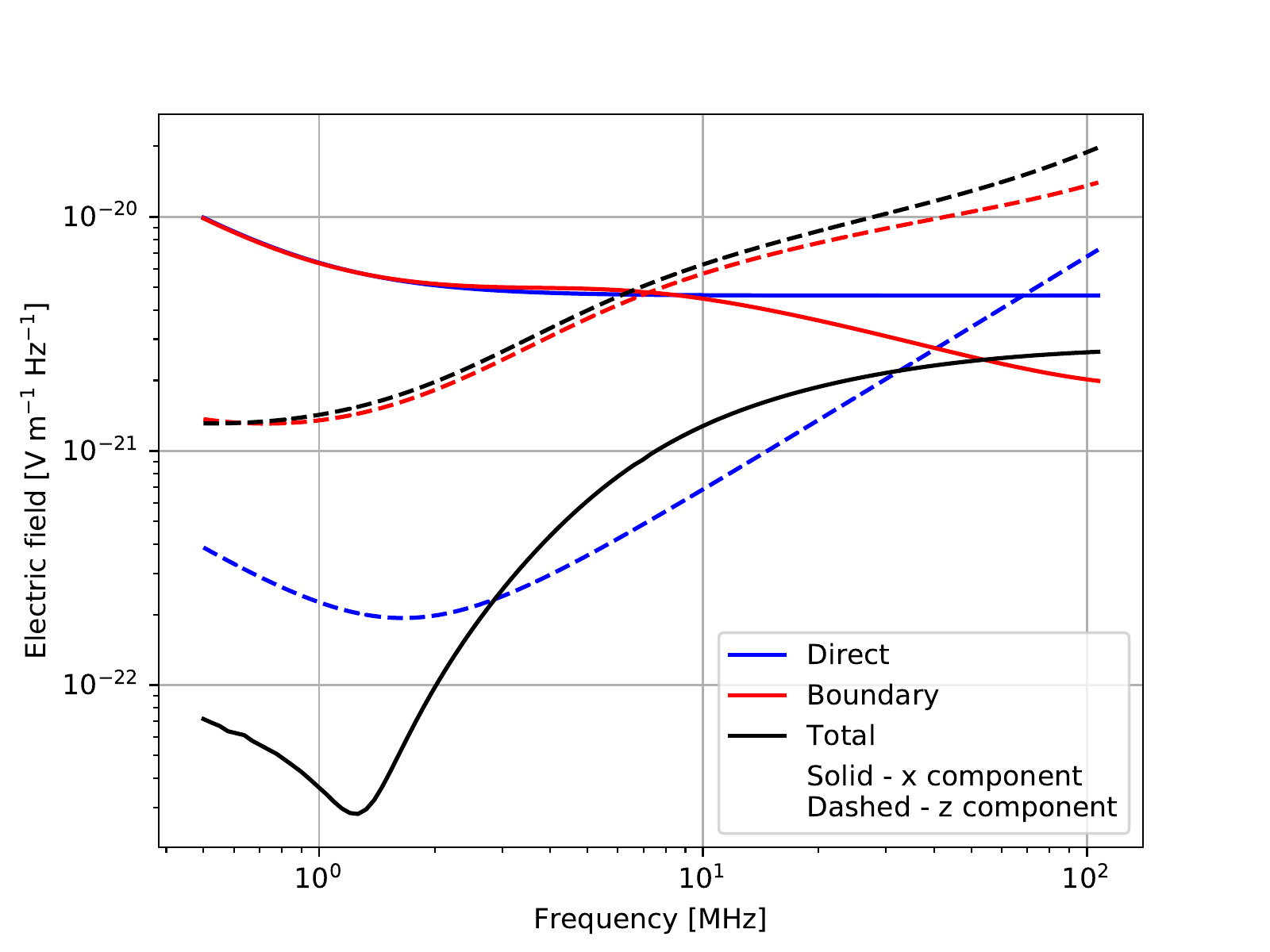}
\caption{Top: Components of the electric field as a function of frequency 
for a $1.2$~m long vertical electron track
traveling at a speed $\sim c$. The end point of the track lies at the boundary.
The vertical ($z$, dashed lines) and horizontal ($x$, solid lines) components are shown.
The direct (blue), boundary (red) and total (black) fields are depicted.
The observer is located at $\rho = 50$~m and $z = 2$~m
Bottom: Same as top, but for a horizontal track at $z = 0.6$~m, 
traveling towards the $+\hat x$ direction.}
\label{fig:trackvert}
\end{figure}

Although Eq.~\eqref{eq:etracksum} for the field of a track in medium 1 conveniently divides
the electric field in direct, image, and integral fields, it is physically more sound to treat the
image and integral fields together as a single entitty, 
since their sum is the field created by the boundary between the two
media and that is what must be added to the direct field created by the track. Symbollycally,
\begin{equation}
\mathbf{E}_\mathrm{track} \xw = 
 \mathbf{E}_\mathrm{track}^{d} + \mathbf{E}_\mathrm{track}^{im} +
\mathbf{E}_\mathrm{track}^{int}
\equiv \mathbf{E}_\mathrm{track}^{d} + \mathbf{E}_\mathrm{track}^{\mathrm{bound}},
\label{eq:etrackbound}
\end{equation}
where $ \mathbf{E}_\mathrm{track}^{\mathrm{bound}}$ represents the field created
by the boundary as a response to the particle track.
Let us place a $1.2$ m long vertical electron track, reaching the ground and stopping at the boundary.
Medium 1 is air and medium 2 is an average soil, as in the previous Section.
In Fig.~\ref{fig:trackvert}, top, we find the field seen by an observer located at $\rho = 50$~m
and $z = 2$~m. The field created by the boundary is quite important. The $x$ component
(solid lines) is completely
dominated by the boundary field, while for the $z$ component
(dashed lines) there is an interplay between the
direct and boundary field, whose interference gives rise to the final form of the total field.
In Fig.~\ref{fig:trackvert}, bottom, we show the field created by a horizontal track at $z = 2$~m.
In this case, the $x$ component (solid lines) is suppressed at low frequencies by the boundary,
while the $z$ component (dashed lines) is boosted.

In any case, the conclusion drawn from Fig.~\ref{fig:trackvert} is clear. A particle track near the
ground creates a non-zero electric field at the position of an observer that lies near ground level also.
Moreover, the boundary field plays an important role, which means that for obtaining a rigorous
field for particles near a boundary, the direct field is not enough.

\begin{figure}
\includegraphics[width=0.7\textwidth]{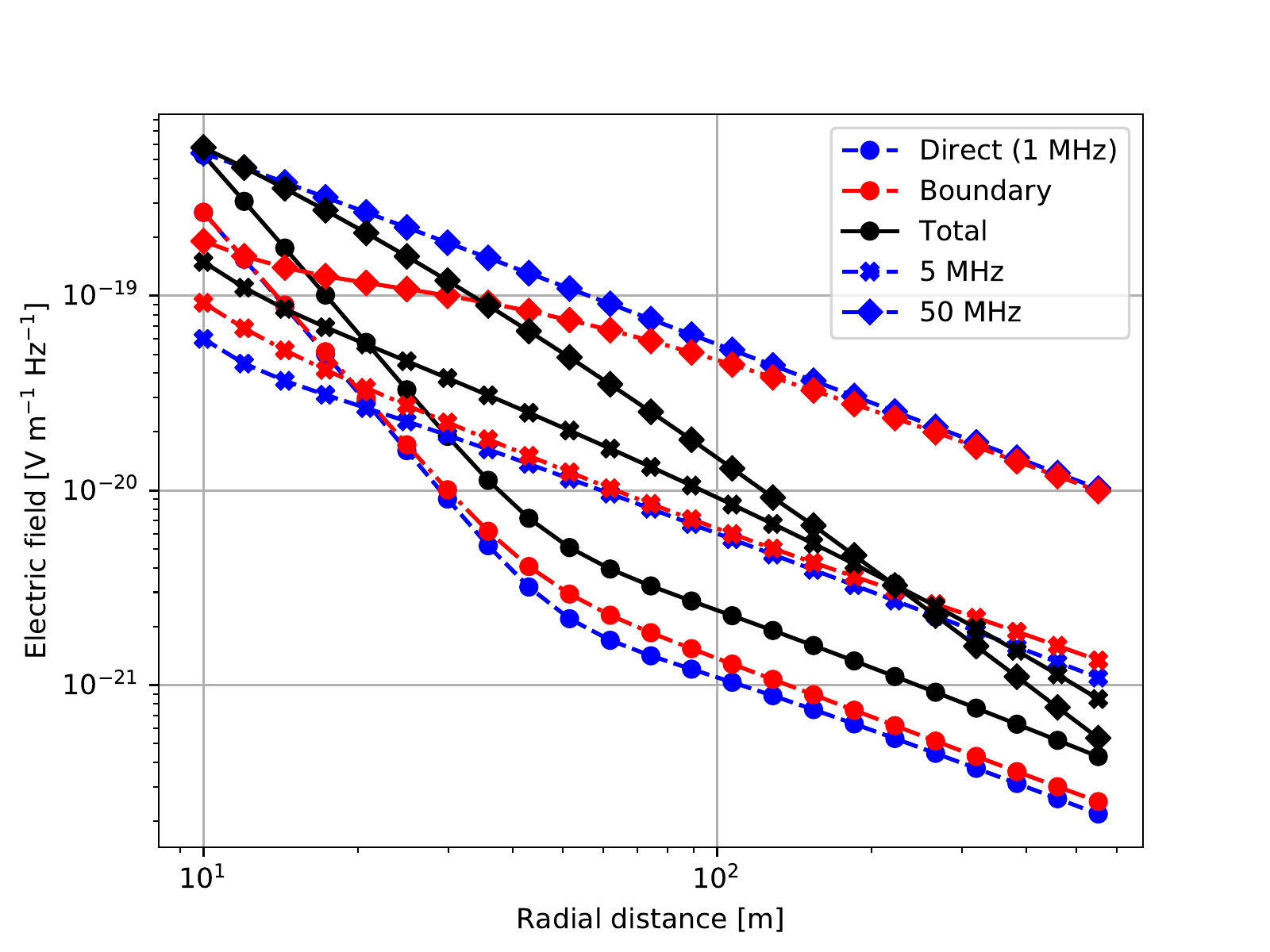}
\caption{Module of the electric field as a function of radial observer distance 
for a $1.2$~m long vertical electron track
traveling at a speed $\sim c$. The end point of the track lies at the boundary.
The direct (blue, dashed lines), boundary (red, dash-dotted lines) and total (black, solid lines) 
fields are depicted, for the frequencies of 1~MHz (circles), 5~MHz (crosses) and
50~MHz (diamonds).
The observers are located at $z = 2$~m.}
\label{fig:trackdistance}
\end{figure}

We show in Fig.~\ref{fig:trackdistance} the electric field for a vertical track near the boundary
as a function of the radial distance to the observer. At 1~MHz and 5~MHz, the boundary amplifies
the emission of the track. However, at 50~MHz, the direct field and the boundary field interfere
destructively, and at several hundreds of meters the field is quite attenuated. Since 50~MHz
is near the middle of the bands usually employed by ground-based cosmic ray detection,
Fig.~\ref{fig:trackdistance} implies that we should expect less emission from the shower
particles near ground level at standard frequencies. 
However, at frequencies below 5~MHz, where most of the 
EXTASIS \cite{antony} band is contained, the total field for a track is greater than the direct field,
and the coherence at these frequencies between different parts of the shower suggests
that the emission from the ground particles of a shower will be amplified thanks to the
boundary as well.

\subsection{Field from an underground track}

\begin{figure}
\includegraphics[width=0.7\textwidth]{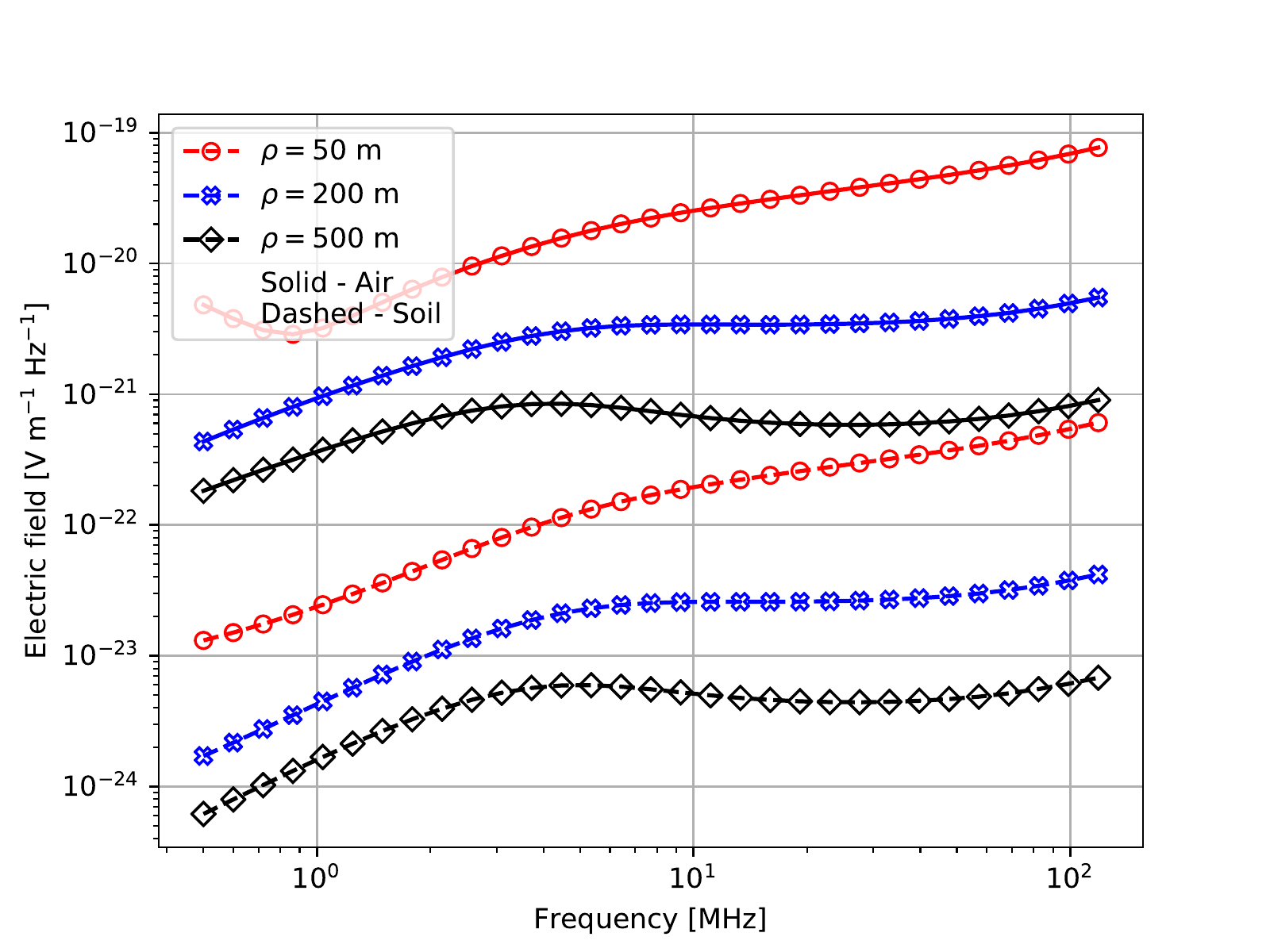}
\caption{Module of the electric field as a function of frequency for an electron track
traveling at a speed $\sim c$ towards the ground and forming a $\pi/4$ angle with the interface.
The track covers $1.2$~m in air, crosses the boundary, and travels $0.12$ m in an average
soil ($\epsilon_r = 12$, $\sigma = 5$~mS/m).
The emission in air (solid lines) and in soil (dashed lines) for observers at radial distances of 
50 (circles), 200 (crosses), and 500~m (diamonds) are shown.
The observers are located at $z = 2$~m. See text for details.}
\label{fig:trackunder}
\end{figure}

Using Eq.~\eqref{eq:etracksumunder} along with Eqs.~\eqref{eq:Evunder} and \eqref{eq:Ehunder},
we can calculate the field from a particle track immersed in soil, which is useful to assess how
the emission in air from the ground particles of the shower relate to the emission in soil, after
they reach the ground. We will consider to that effect a vertical electron track trat travels
$1.2$~m in air with an angle of $\pi/4$ with respect to the boundary, reaches the ground, 
and then travels another $0.12$~m in soil in the same direction. We expect the underground
particles for a particle shower to be stopped after a few centimeters, so the length of the
underground track is reasonable. We show in Fig.~\ref{fig:trackunder} the electric fields created
by the track, both in air and in soil. We see that the field emitted underground is two orders
of magnitude lower than the field emitted in air. We have checked that for vertical tracks,
this difference is slightly larger. Therefore, for downward-going tracks, we can ignore the
field from the underground track and calculate only the emission when the track is in the atmosphere.

\subsection{Influence of observer height on the electric field from a track}

\begin{figure}
\includegraphics[width=0.7\textwidth]{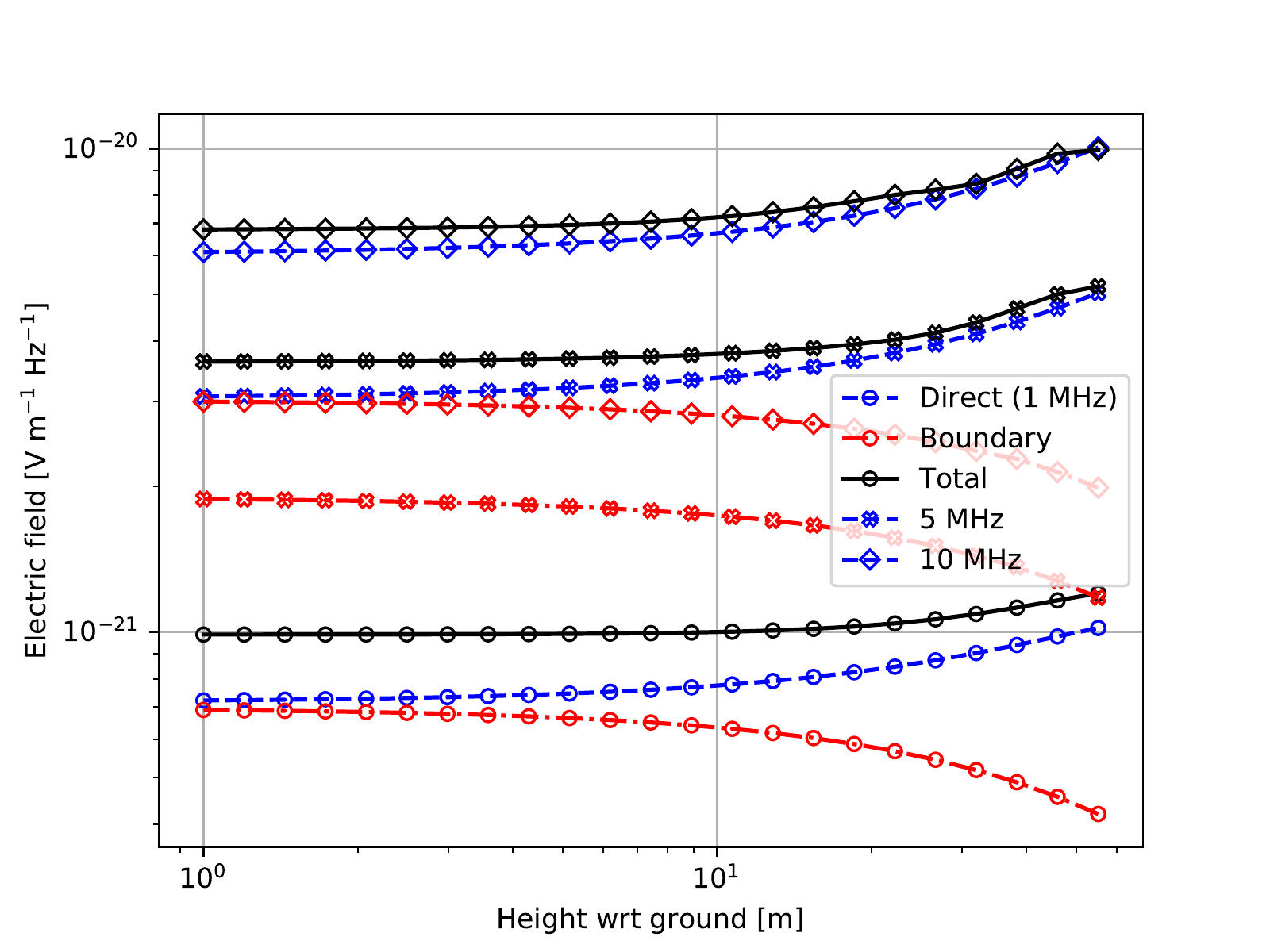}
\caption{Module of the electric field as a function of observer height for a 
$1.2$~m long vertical electron track traveling at a speed $\sim c$.
The direct (dashed lines), boundary (dash-dotted lines) and total fields (black lines)
for 1 (circles), 5 (crosses), and 10~MHz (diamonds) are shown.
The observers are located at $\rho = 100$~m. See text for details.}
\label{fig:trackheight}
\end{figure}

The field created by the boundary depends on the observer's position with respect to it.
As a result, the height of the observer will influence the total field seen by the observer.
To illustrate this influence, we have taken a vertical track with $z' = 100$~m over a ground
with $\epsilon_r = 12$ and $\sigma = 5$~mS/m,
and we have calculated the field for a series of observers located at $\rho = 100$~m and
variable height. The results are shown in Fig.~\ref{fig:trackheight}. We show the fields
for 1, 5, and 10~MHz. At 1~MHz, the dependence of the total with height is less
pronounced than at 5 and 10~MHz. 
At 10~MHz, the observation wavelength ($\sim 30$~m) is about one third of the track's
height (100~m), and since the ground lies near the far-field region of the track, the total
field presents an interference pattern similar to the one created by the sum of a direct and a reflected field.
We explain in the next section that this is indeed the case.

\subsection{Far field. Decomposition into direct, reflected, and transmitted components}
\label{sec:fresnel}

\begin{figure}
\includegraphics[width=0.7\textwidth]{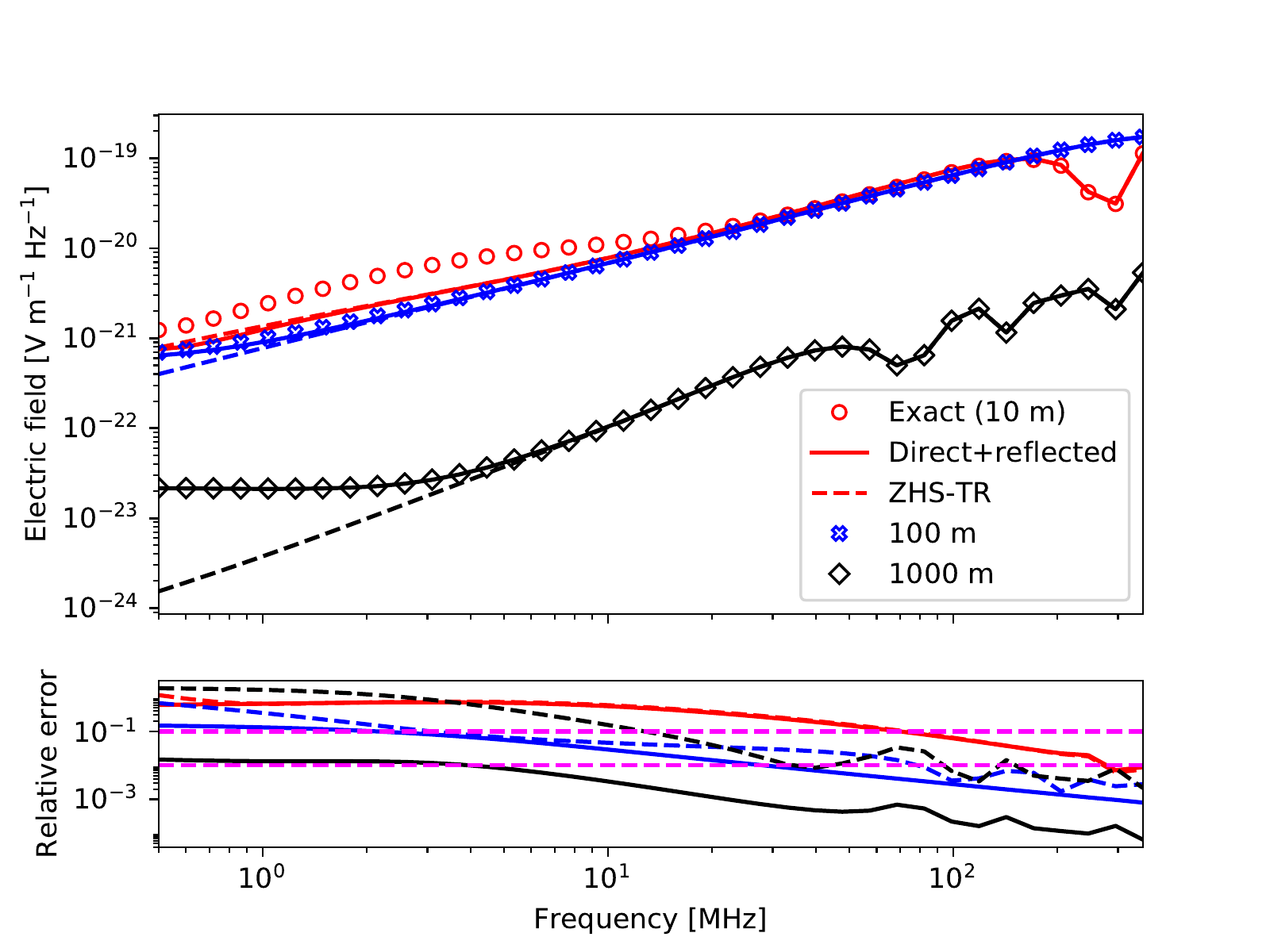}
\includegraphics[width=0.7\textwidth]{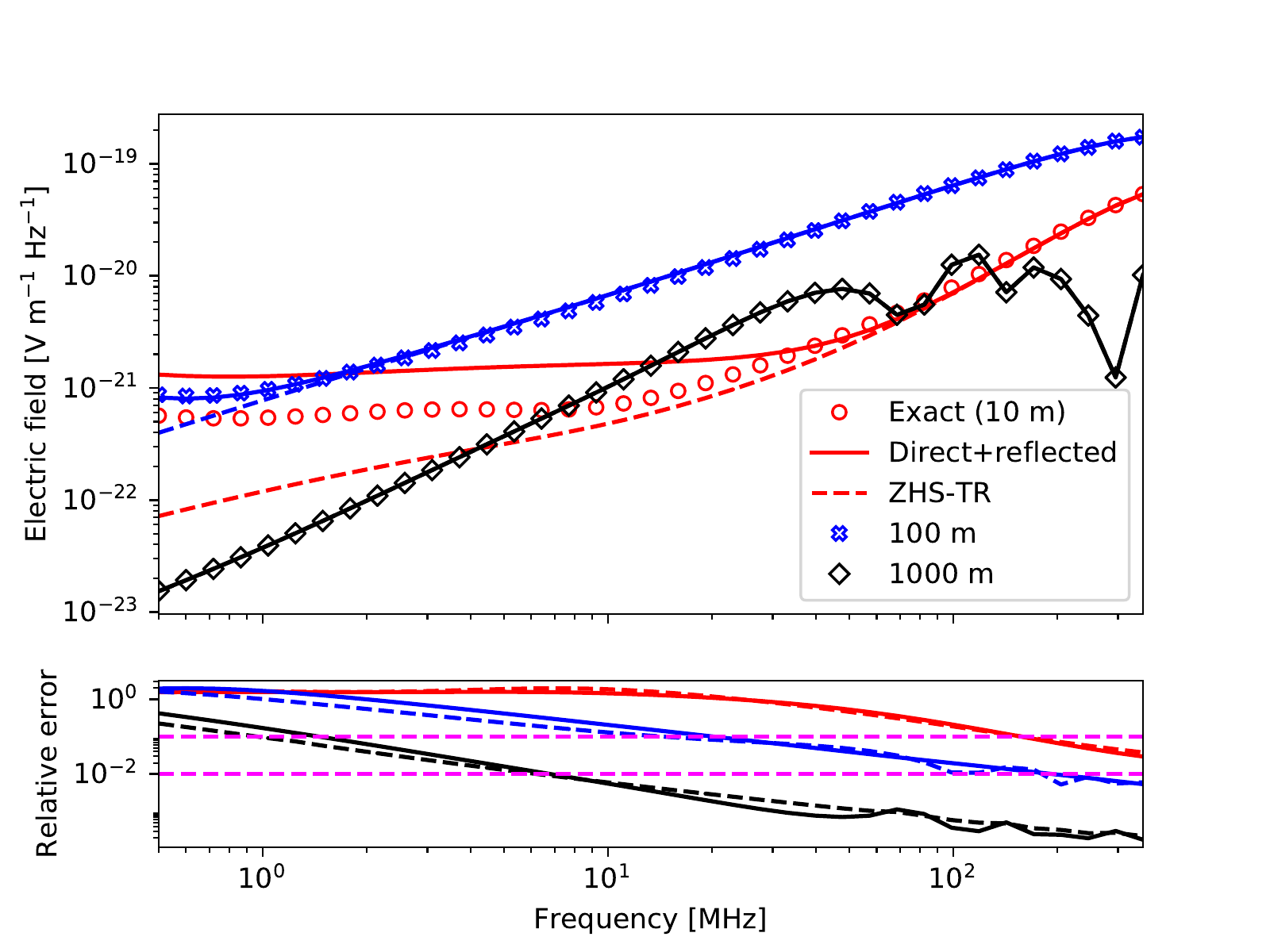}
\caption{The four figures are called A, B, C, and D from top to bottom.
A: Electric field as a function of frequency created by a $1.2$~m long electron track
traveling at a speed $\sim c$. The middle point of the track is located at 10,
100 and 1000~m above the interface
and the track is traveling downwards ($-\hat z$) towards the boundary. 
The observer is located at $\rho= 100$~m and $z = 10$~m.
Medium 1 is air and medium 2 is an average soil.
The exact fields, the Fresnel (direct plus reflected) approximation and the ZHS-TR formula
are shown. B: relative errors of the Fresnel and ZHS-TR approaches with respect to
the exact formula. The dashed magenta line indicates an error of $0.01$ ($1\%$).
C and D: Same as A and B, but for a horizontal ($+\hat x$) track.
See text for details.}
\label{fig:fresnelup}
\end{figure}

If either the emitting track or the observer are far away from the boundary compared to
the wavelength ($z$ or $z' \ll \lambda$), the field should reduce to a sum of direct and
reflected fields if the track is in medium 1. With the help of Fresnel coefficients, and using
the direct (Eq.~\eqref{eq:edfraun}) and image (Eq.~\eqref{eq:eimfraun}) fields, the
far-field approximation can be written as:
\begin{equation}
\mathbf{E}^\mathrm{far}_\mathrm{track} = \mathbf{E}^\mathrm{d}_\mathrm{track} 
+ r_\parallel \mathbf{E}^\mathrm{im}_\mathrm{track,\parallel}
+ r_\perp \mathbf{E}^\mathrm{im}_\mathrm{track,\perp}.
\label{eq:Ereflected}
\end{equation}
The image field has been divided into the polarizations perpendicular and parallel 
to the reflection plane.
The parallel ($r_{\parallel}$) and perpendicular ($r_{\perp}$) Fresnel reflection coefficients
can be written with the help of the following equations, knowing the complex wavenumbers 
$k_1$ and $k_2$ for each medium and the reflection angle $\theta_1$:
\begin{equation}
\Re(k_2) \sin\theta_2 = \Re(k_1) \sin\theta_1; \qquad
\alpha \equiv \frac{\cos\theta_2}{\cos\theta_1}; \qquad
\beta \equiv \frac{k_2}{k_1}
\label{eq:alphabeta}
\end{equation}
\begin{equation}
r_\parallel  =  \frac{\alpha-\beta}{\alpha+\beta}; \qquad
r_\perp =  \frac{1 - \alpha\beta}{1 + \alpha\beta}
\label{eq:fresnelup}
\end{equation}
Note that $\theta_2$ is the transmitted angle, given by Snell's law. We can compare now
the exact calculation (Eq.~\eqref{eq:etracksum}) with the Fresnel approximation 
(Eq.~\eqref{eq:Ereflected}). We will also compare with the ZHS-TR method \cite{motloch},
that combines the ZHS formula and the Fresnel coefficients. If we keep only the leading terms
 that fall with $\frac{1}{r}$ in Eqs.~\eqref{eq:Evd}, \eqref{eq:Ehd}, \eqref{eq:Evim}, and
 \eqref{eq:Ehim}, we arrive at the same formula. In \cite{prd87} it is proven
 that the direct field of a vertical track reduces to the ZHS formula in the far field. 
 Since the field from a horizontal track can be obtained upon rotation
 (Appendix~\ref{sec:vtoh}),  the field from a horizontal track reduces as well to the 
 ZHS formula in the far field. The image field, and therefore the reflected field, 
 has the same functional dependence (compare Eqs.~\eqref{eq:Evd} and \eqref{eq:Ehd} to
 Eqs.~\eqref{eq:Evim} and \eqref{eq:Ehim}), which implies that
 it reduces as well to the ZHS formula in the far field. Adding the Fresnel coefficients
to the direct and image ZHS fields $\mathbf{E}_\mathrm{ZHS}$, we retrieve the ZHS-TR method:
 \begin{equation}
 \mathbf{E}^\mathrm{far}_\mathrm{ZHS} = \mathbf{E}^\mathrm{d}_\mathrm{ZHS} 
+ r_\parallel \mathbf{E}^\mathrm{im}_\mathrm{ZHS,\parallel}
+ r_\perp \mathbf{E}^\mathrm{im}_\mathrm{ZHS,\perp}
 \end{equation}
We show in Fig.~\ref{fig:fresnelup} a comparison between the exact calculation (Eq.~\eqref{eq:etracksum}),
the Fresnel approximation (Eq.~\eqref{eq:Ereflected}) and the ZHS-TR formula for a downward going
vertical electron track (top) and a horizontal electron track (bottom) at three heights:
10, 100 and 1000~m. We have placed an observer, at a radial distance of 100~m and 10~m of height.
Fig.~\ref{fig:fresnelup} shows that the exact approach, the Fresnel approximation and the ZHS-TR
method agree in the far field. However, for low enough frequencies, neither the Fresnel approximation
nor the ZHS-TR method agree with the field predicted by the exact method. When the observation
wavelength is 3 times smaller than the distance between the track and the interface, the relative
error is around $10\%$. 
\begin{equation}
\frac{z'}{\lambda} \gtrsim 3 \qquad \Rightarrow \qquad \mathrm{error} \lesssim 10\%
\end{equation}
Putting it in terms of the frequency, for air:
\begin{equation}
\frac{\nu}{1\mathrm{~MHz}} \frac{z'}{1\mathrm{~km}}
\gtrsim 1 \qquad \Rightarrow \qquad \mathrm{error} \lesssim 10\%
\end{equation}

We can also verify that the field emitted by a track in medium 2 can be expressed in the far field
as a transmitted field with the help of the Fresnel coefficients. In this case, we are going to assume
that medium 1 is air as before, and medium 2 is lossless Antarctic ice with a refractivity of
$n_\mathrm{ice} = 1.78045$. Several observers are placed at a radial distance of $100$~m
and heights of 10, 100 and 100~m. This physical configuration is similar to the setup for the
ARIANNA \cite{arianna} antennas that are located on snow, although with larger
heights\footnote{Note that the present
work's formalism can also be applied to study the influence of the air/ice interface when the
antennas and the emitted particles are embedded in ice. It suffices to identify medium 1 as
ice, place the antennas and particles there, and identify medium 2 as air.}.

\begin{figure}
\includegraphics[width=0.7\textwidth]{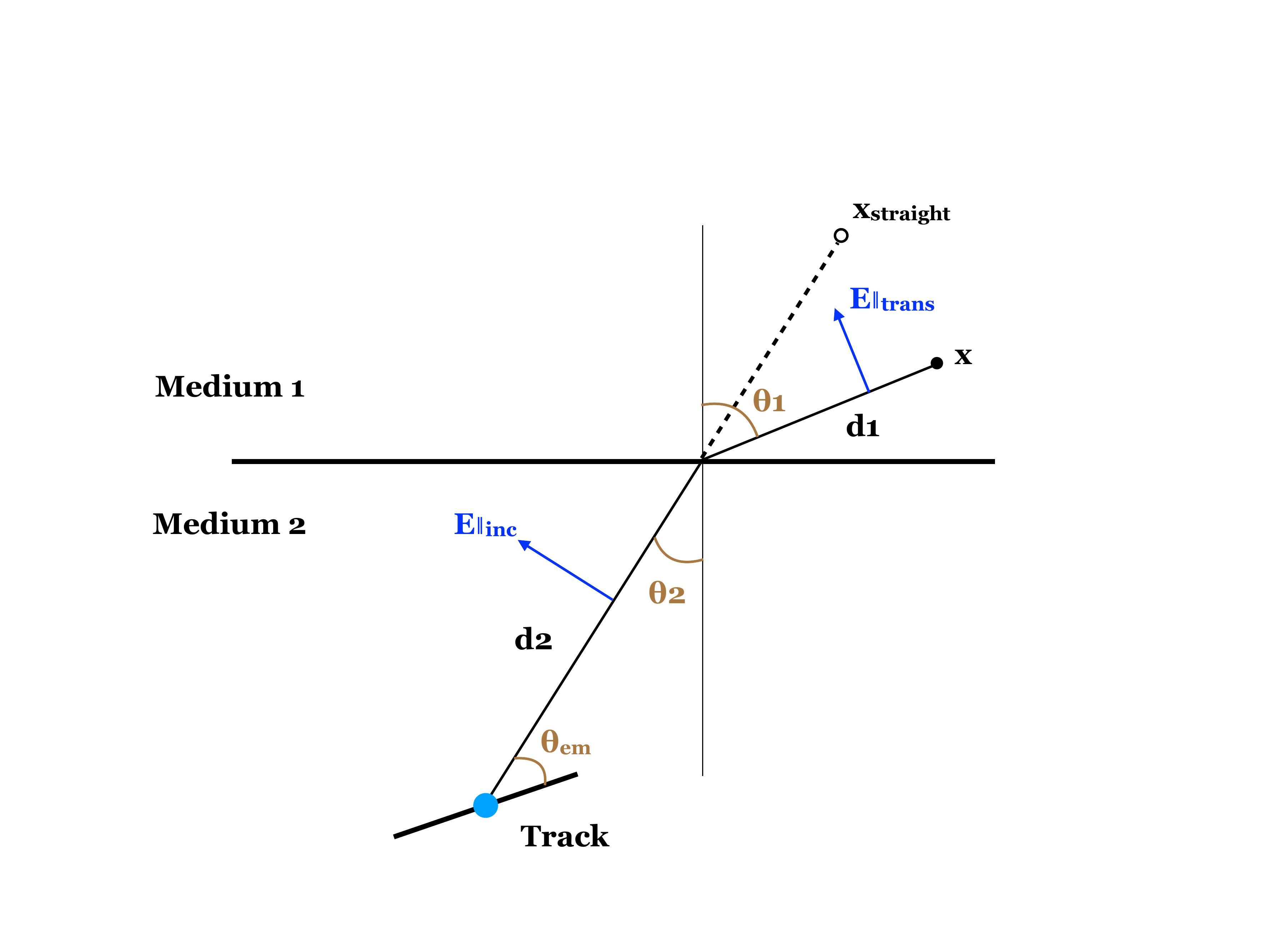}
\caption{Sketch explaining the transmission geometry. 
The figure depicts the refraction plane. The track lies in medium 2, emitting an
electromagnetic wave with an emission angle $\theta_\mathrm{em}$. The wave field parallel
to the refraction plane ($\mathbf{E}_{\parallel\mathrm{inc}}$) arrives at the boundary, and its
direction of propagation forms an angle $\theta_2$ with the vector normal to the boundary.
The field is refracted and enters medium 1 forming a $\theta_1$ angle with the
normal to the boundary. As a consequence, the transmitted field 
$\mathbf{E}_{\perp\mathrm{trans}}$ is rotated with respect to the incident field. The transmitted
field then arrives at the observer located in $\mathbf{x}$. 
$d_{2(1)}$ is the distance traveled by the wave in medium 2(1).
$\mathbf{x}_\mathrm{straight}$ denotes the apparent position of $\mathbf{x}$ seen
from the track position.}
\label{fig:sketch}
\end{figure}

The exact field $\mathbf{E}_{\mathrm{track},2\rightarrow 1}$ when the particle is in medium
2 and the observer in medium 1can be approximated in the far field
using the direct field for medium 2 (we suppose that there is no medium 1), 
the Fresnel transmission coeffcients, a rotation matrix, and a phase to correct
for the actual path taken by the wave:
\begin{equation}
\mathbf{E}^\mathrm{far}_{\mathrm{track},2\rightarrow 1} =
e^{i ( k_1 - k_2 ) d_1} \left[
R(\theta_1-\theta_2) T_\parallel 
\mathbf{E}^\mathrm{d2}_\mathrm{track,\parallel}(\mathbf{x}_\mathrm{straight})
+ T_\perp \mathbf{E}^\mathrm{d2}_\mathrm{track,\perp}(\mathbf{x}_\mathrm{straight}) \right]
\label{eq:fresneldown}
\end{equation}
$d_{1(2)}$ is the distance travelled by the radiation in medium 1 (2). The phase is there
to account for the fact that the direct field in medium 2 does not consider the propagation
in medium 1.
The direct field is calculated for a point named $\mathbf{x}_\mathrm{straight}$, which is the
point that lies at a distance $d_1+d_2$ and along a line that forms a $\theta_2$ angle with the
normal to the boundary\footnote{Or, equivalently, the apparent location of the observer if
we consider the opposite optical path - a ray emitted by the observer that reaches the particle's
position.}. This is necessary to ensure that the radiation dependence goes as $1/(d_1+d_2)$ as intended,
as well as for calculating the correct emission angle for the radiation which gives the correct
polarization. In this case $\theta_2$ is the angle that forms the incident wave with the normal
and $\theta_1$ the refracted angle. We have calculated these angles numerically, since given two
points at each side of the interface, there is no general analytical solution for the incident
and refracted angles. See Fig.~\ref{fig:sketch} for visual help.

The parallel ($T_\parallel$) and perpendicular ($T_\perp$) transmission coefficients are the Fresnel
transmission coefficients multiplied by a correcting factor. Fresnel coefficients are meant to be used
for plane waves, but the refraction that takes place when the wave goes from a dense medium
to a light medium induces
a divergence of the rays, effectively diminisihing the electric field. Following \cite{endpoints,motloch},
we write this factor as:
\begin{equation}
\frac{\mathrm{d}\theta_2}{\mathrm{d}\theta_1} = 
\frac{\Re(k_1)\cos\theta_1}{\Re(k_2)\cos\theta_2},
\end{equation}
which multiplied by the Fresnel transmission coefficients results in:
\begin{equation}
T_\parallel = \frac{2\Re(k_1)\cos\theta_1}{ \Re(k_1)\cos\theta_2 + \Re(k_2)\cos\theta_1 };
\qquad
T_\perp = \frac{ 2\Re(k_1)\cos\theta_1 }{ \Re(k_2)\cos\theta_2 + \Re(k_1)\cos\theta_1 }
\end{equation}

The rotation matrix $R(\theta_1-\theta_2)$ in Eq.~\eqref{eq:fresneldown}
represents a rotation of an angle $\theta_1-\theta_2$
around the vector normal to the refraction plane, and it is due to the fact that the refraction
rotates the polarization of the parallel component of the incident wave. See Fig.~\ref{fig:sketch}.

\begin{figure}
\includegraphics[width=0.7\textwidth]{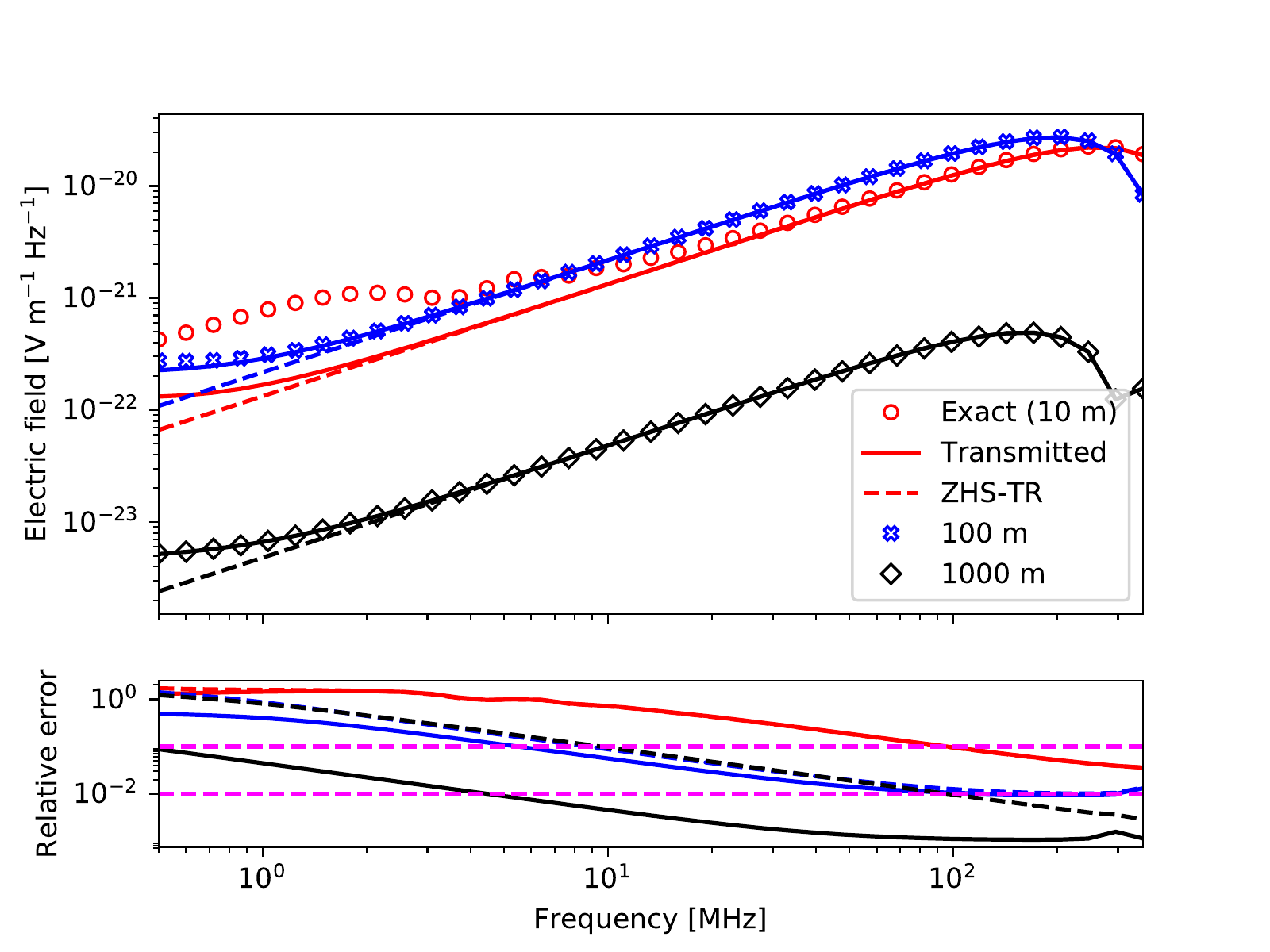}
\includegraphics[width=0.7\textwidth]{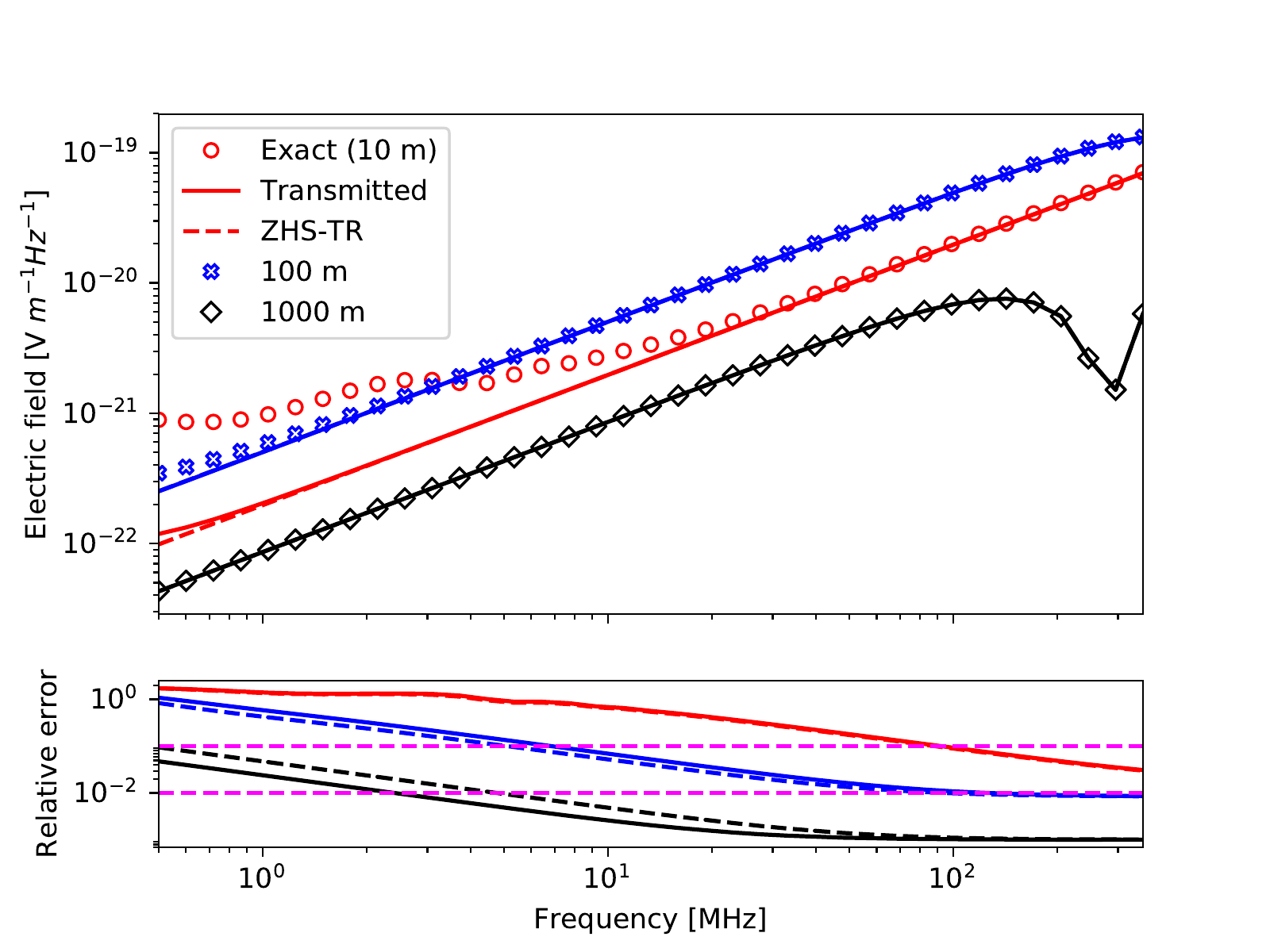}
\caption{The four figures are named A, B, C, and D from top to bottom.
A: Electric field as a function of frequency created by a $1.2$~m long electron track
traveling at a speed $\sim c$. The middle point of the track is located at $z = -2$~m, below
the interface (medium 2)
and the track is traveling towards the boundary with an angle of $\pi/4$ with respect to
the normal. The observers are located at $\rho= 100$~m and $z = 10$~m, 100, and 1000~m.
The exact fields, the Fresnel (transmitted) approximation and the ZHS-TR formula
are shown. B: relative errors of the Fresnel and ZHS-TR approaches with respect to
the exact formula. The dashed magenta line indicates an error of $0.01$ ($1\%$).
See text for details.}
\label{fig:fresneldown}
\end{figure}

Let us now consider an electron track in ice (medium 2) whose middle point is at $z' = -2$. 
The track is going upwards, forming a $\pi/4$ angle
with the normal to the boundary. We place several observers at $\rho = 100$~m and
$z = 10$, 100, and 1000~m. As we can see in Fig.~\ref{fig:fresneldown}, the exact, Fresnel
and ZHS-TR calculations agree when frequency is high enough, as it happened with
Fig.~\ref{fig:fresnelup}. In this case what changes is the distance from the boundary
to the observer (and not the track), but the criterion for the relative error remains unchanged
nonetheless. When the $z$ coordinate is 3 times greater than the wavelength, the error
is less than $10\%$.
\begin{equation}
\frac{z}{\lambda} \gtrsim 3 \qquad \Rightarrow \qquad \mathrm{error} \lesssim 10\%
\end{equation}

We can conclude that if the particles or the antennas are at a distance from boundary three times larger than
the observation wavelength, the decomposition into direct, reflected and transmitted components coupled
to the Fresnel coefficients is a good approximation ($\sim 10\%$ of accuracy) 
to the exact field. If it is not the case,
the exact approach should be used. These results validate the far-field approaches taken
in \cite{motloch,endpoints}.

\section{Shower toy model. Sudden death pulse}
\label{sec:model}

After developping and studying the field from a particle track in Section~\ref{sec:track}, it
would be desirable to implement the formulas in a Monte Carlo code like SELFAS in order to
obtain an accurate prediction for the influence of the ground on the electric field emitted
by an EAS. However, the numerical integration of Eqs.~\eqref{eq:Evint} and \eqref{eq:Ehint}
requires an important CPU time. Even applying the Fresnel approximation when the track
is far enough from the ground (as explained in Section~\ref{sec:fresnel}), the computation
of the field of a track under 20~MHz takes $\sim 10$~s for a single antenna, since each
particle trajectory is made up of several tracks, and the field must be known for a set of
frequencies. A shower comprised of ten million particles (whether they are 
sampled as in SELFAS or thinned as in ZHAireS and CoREAS) would take $10^{8}$~s,
more than 3~years.

As an alternative, we will undertake the calculation of the field emitted by a simple model
for an EAS. The aim of this model is to elucidate the modification of the direct field induced
by the interface, which constitues a correction to the direct field prediction shown 
in \cite{sdparxiv}, and also to verify if the sudden death pulse induced by the
coherent deceleration of the shower front is still present.

Let us model the shower as a collection of straight lines stretching from an altitude of
10~km to the ground level. These lines are positioned at the following radial distances
from the shower core: $\rho = \{10, 30, 50, 70, 90\}$~m. For each one of these distances,
we place eight vertical lines at the azimuthal angles $\varphi = \{0,\pi/4, \pi, \dots, 7\pi/4\}$. 
In total, our toy model is made up of 40 long lines. Each line is divided into 3~m long particle
tracks whose charge depends on the height $z$ and the radial distance to the shower core:
\begin{equation}
q_i = -0.2 N(z_i) f(\rho_i) A_i,
\end{equation}
$N(z_i)$ is the number of particles at the height $z_i$, taken from a Gaisser-Hillas distribution
for a 1~EeV shower in our case. The factor $-0.2$ comes from the negative excess charge in the shower.
$f(\rho_i)$ represents the lateral distribution of
the shower particles, modeled by a Nishimura-Kamata-Greisen (NKG) function. Since our model
effectively transforms the shower into a set of one-dimensional subshowers
each one of these subshowers approximate a portion of the shower front with
an area given by
\begin{equation}
A_i = \frac{\pi}{4} \left( (\rho_i + \frac{\Delta\rho}{2})^2
- (\rho_i - \frac{\Delta\rho}{2})^2 \right),
\end{equation}
with $\Delta \rho = 20$~m.

The 80 subshowers start developping at $z = 10$~km, traveling at $v \sim c$. Every 3~m,
the tracks are stopped and another track having a different charge emerges from the stopping
point. This process is repeated until the whole shower arrives at ground level, $z_g$. The field
is calculated for a given observer and several frequencies under $20$~MHz, where the boundary
effects are going to be more prominent. If the height of the track is greater than the wavelength
by a factor of 10, the field is calculated using the Fresnel approximations 
(Eq.~\eqref{eq:Ereflected}). Otherwise, the exact field is calculated by means of
Eq.~\eqref{eq:etrack}. Once the fields for the chosen set of frequencies has been calculated,
the data are filtered using an eight-order low-pass Butterworth filter 
(with $\nu_c = 10$~MHz as critical frequency) zero-padded, and transformed
to time domain to obtain the time trace of the electric field.

\begin{figure}
\includegraphics[width=0.7\textwidth]{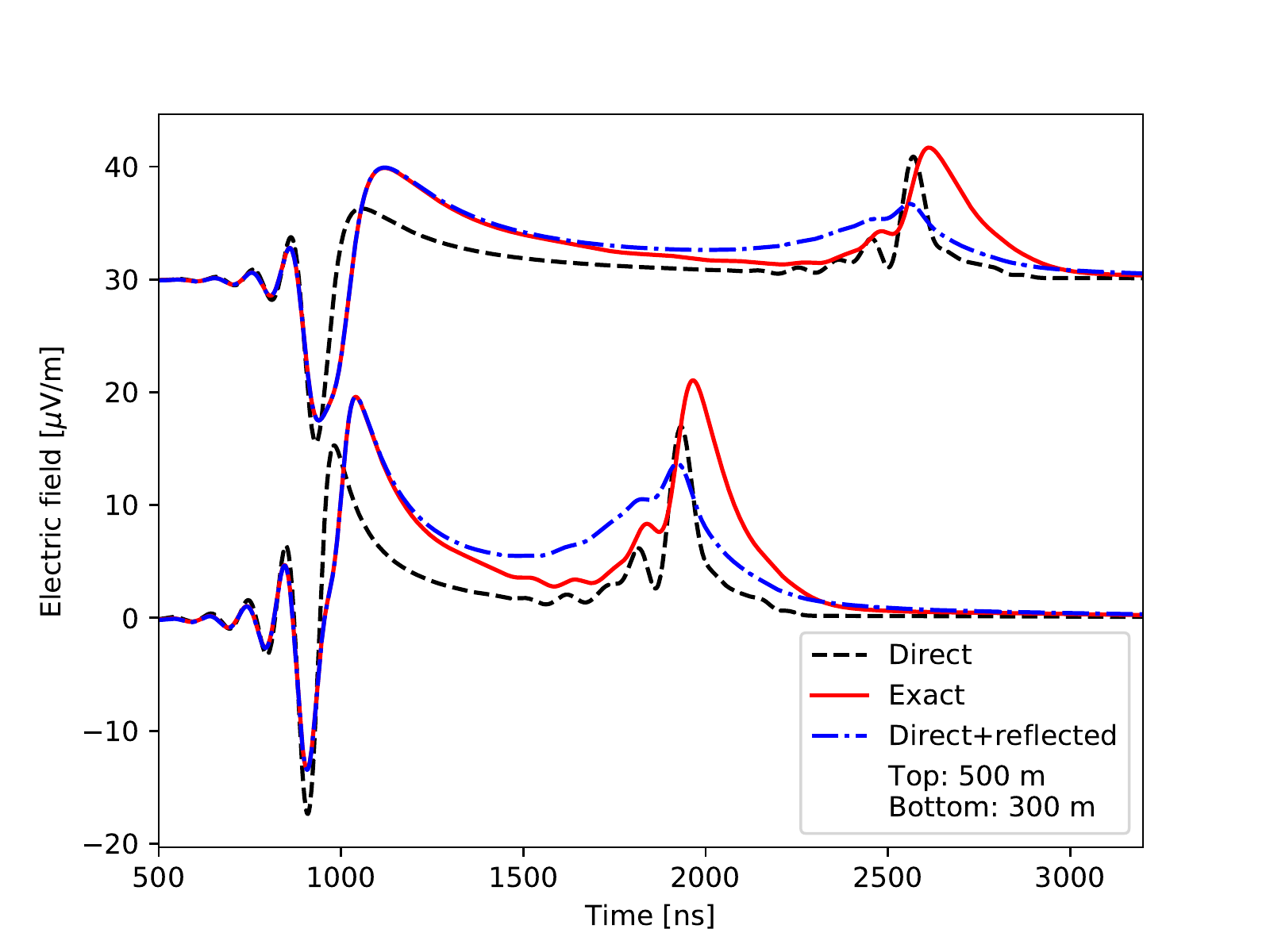}
\caption{Electric field in time domain created by our model shower.
traveling at a speed $\sim c$. 
Medium 1 is air and medium 2 is an average soil.
The observers are located at $\rho= 300$ (lower curves) and $500$~m 
(upper curves, electric field offset by $30$~$\mu$V/m) and $z = 9$~m.
The exact fields (solid red line), the Fresnel approximation (direct+reflected,
dash-dotted line) and the direct (dashed black lines) are shown.
See text for details.}
\label{fig:modeltime}
\end{figure}

\begin{figure}
\includegraphics[width=0.7\textwidth]{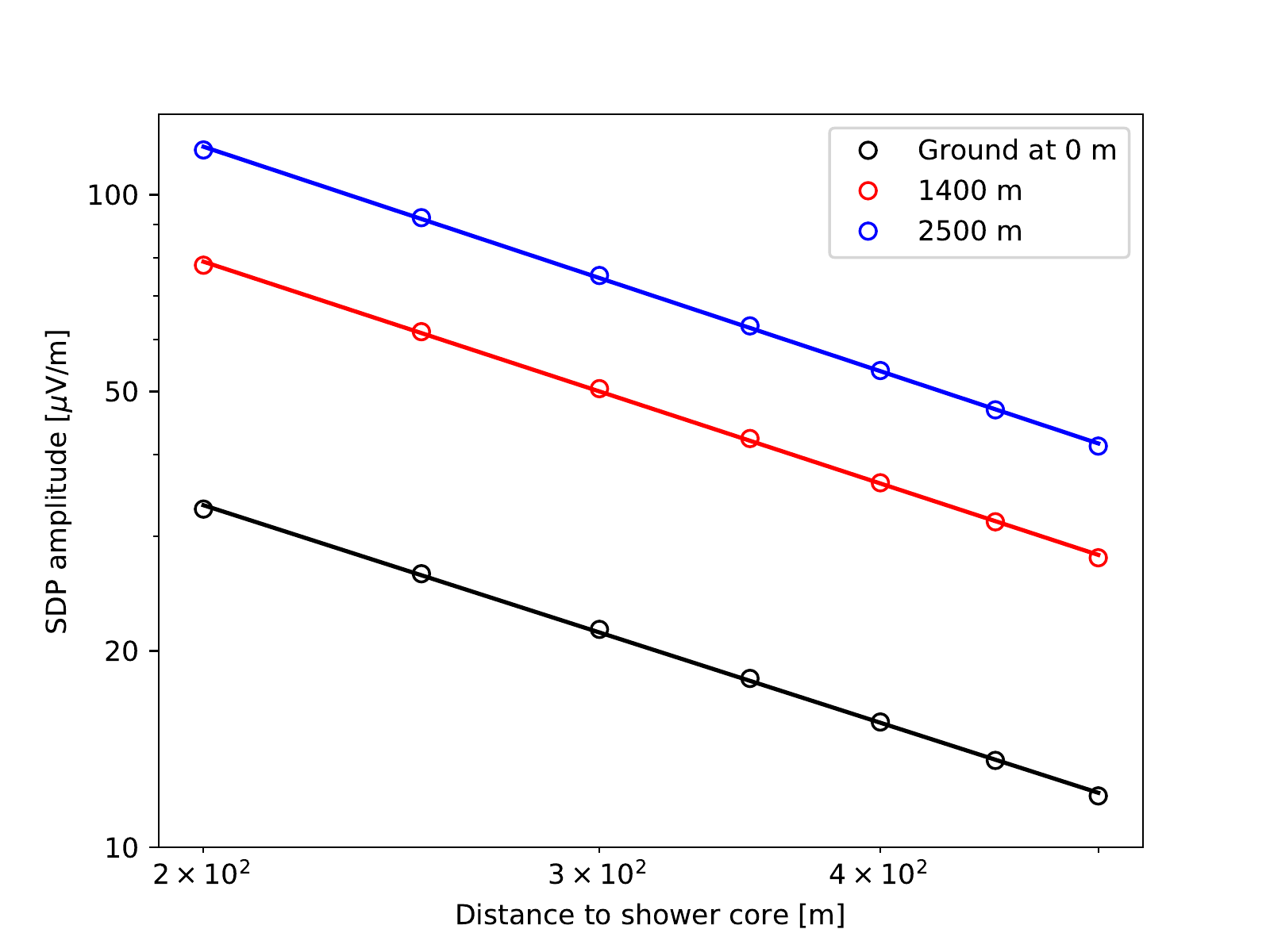}
\includegraphics[width=0.7\textwidth]{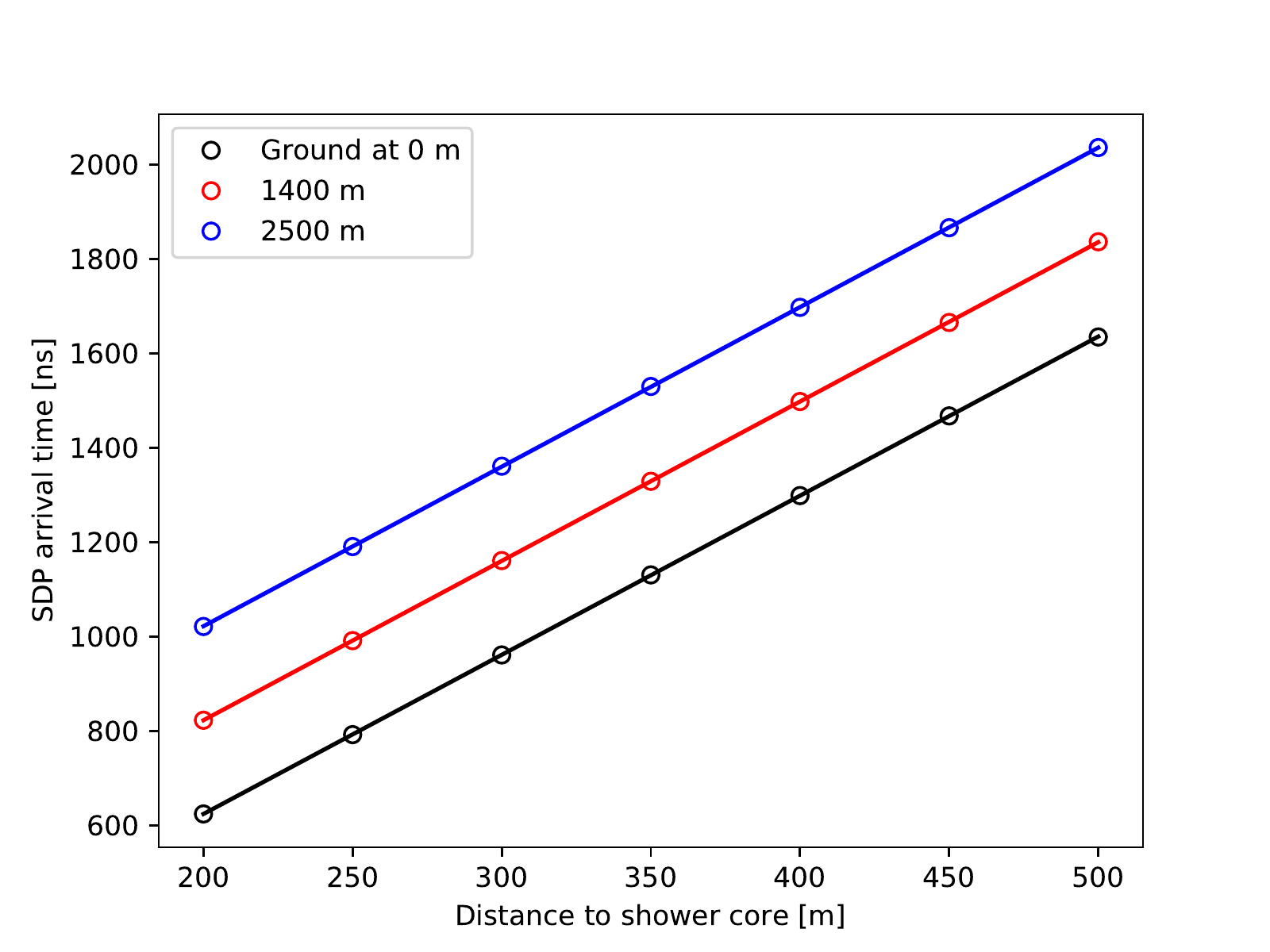}
\caption{Top: Amplitude of the SDP maximum created by a vertical shower as function of the distance
to the shower core. The observers are located at a height of 2~m above the ground.
Ground altitudes are 0~m (lower curve), 1400~m (middle curve) and 2500~m (upper curve).
Points represent the SDP amplitudes and lines show a $1/R$ fit to the data, with $R$ being
the distance from the observer to the shower core.
Medium 1 is air and medium 2 is an average soil.
Bottom: Same setup as top, but now the $y$ axis shows the arrival time of the SDP maximum,
zero being the time when the shower reaches the ground (data for 1400~m and 2500~m are
offset by 200 and 400~ns, respectively). Lines indicate a linear fit for the arrival time.
See text for details.}
\label{fig:disttime}
\end{figure}

We can find in Fig.~\ref{fig:modeltime} the electric field in time domain given by our model for
a vertical shower. Ground altitude is $z_g = 0$~m, and we have placed two observers at
$300$ and $500$~m of radial distance and at 9~m of height, which is the height of
the antennas used for the EXTASIS experiment. Medium 1 is air, with an
 $\epsilon_r = (1.0001)^2$ and no conductivity, and medium 2 is
again an average soil with $\epsilon_r = 12$ and $\sigma = 5$~mS/m.
The direct (Eq.~\eqref{eq:edfraun}, dashed black
lines), direct plus reflected (Fresnel approximation, Eq.~\eqref{eq:Ereflected}, dash-dotted blue
lines) and exact (Eq.~\eqref{eq:etrack}, solid red lines) are plotted. The left part of the trace is due to the
shower maximum. Since the maximum is located at several kilometers of altitude, the Fresnel
approximation and the exact calculation agree on the emission from the shower maximum.
As time passes and the shower develops, the particles get closer to the ground and the
Fresnel approximation ceases to be valid. When the particles stop at ground level, their
sudden deceleration creates the second peak that can be seen around $2000$~ns for
the observer at $300$~m (bottom) and aroud $2500$~ns for the observer at $500$~m
(top). This is the sudden death pulse (SDP) already discussed in \cite{sdp,sdparxiv},
although in those references only the direct emission had been computed. The present work's
approach shows that although the exact calculation and the direct emission differ, the SDP
is however still present and presents similar properties to the ones outlined by the direct calculation:
\begin{enumerate}
\item The amplitudes and durations of the exact and direct SDPs are of the same order of magnitude.
\item The delay between the principal pulse and the SDP is directly proportional to the distance
from the shower core to the observer (see Fig.~\ref{fig:disttime}, bottom), as with the direct case.
\item The amplitude of the SDP falls with the inverse of the distance from shower core to observer
(see Fig.~\ref{fig:disttime}, top), as it happened with the direct case.
\end{enumerate}
One interesting feature found in Fig.~\ref{fig:modeltime} is the different time at with the maximum
of the SDP occurs for the direct (or direct+reflected) and the exact calculation. The exact maximum
arrives slightly later than the direct or direct+reflected maximum, and the reason for that is the
surface wave (or lateral wave) that propagates along the boundary and therefore has a longer path
than the direct and reflected waves.

\begin{figure}
\includegraphics[width=0.49\textwidth]{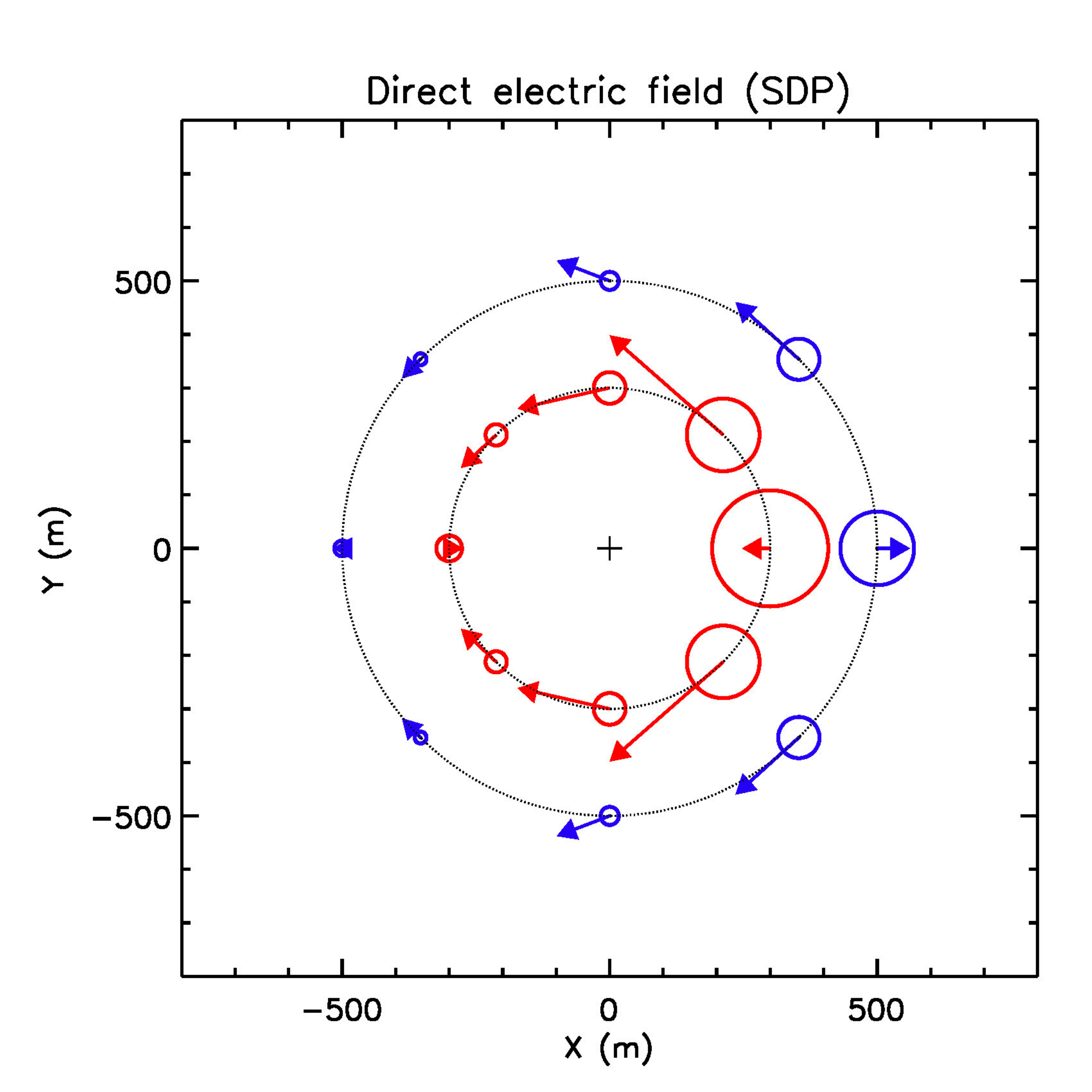}
\includegraphics[width=0.49\textwidth]{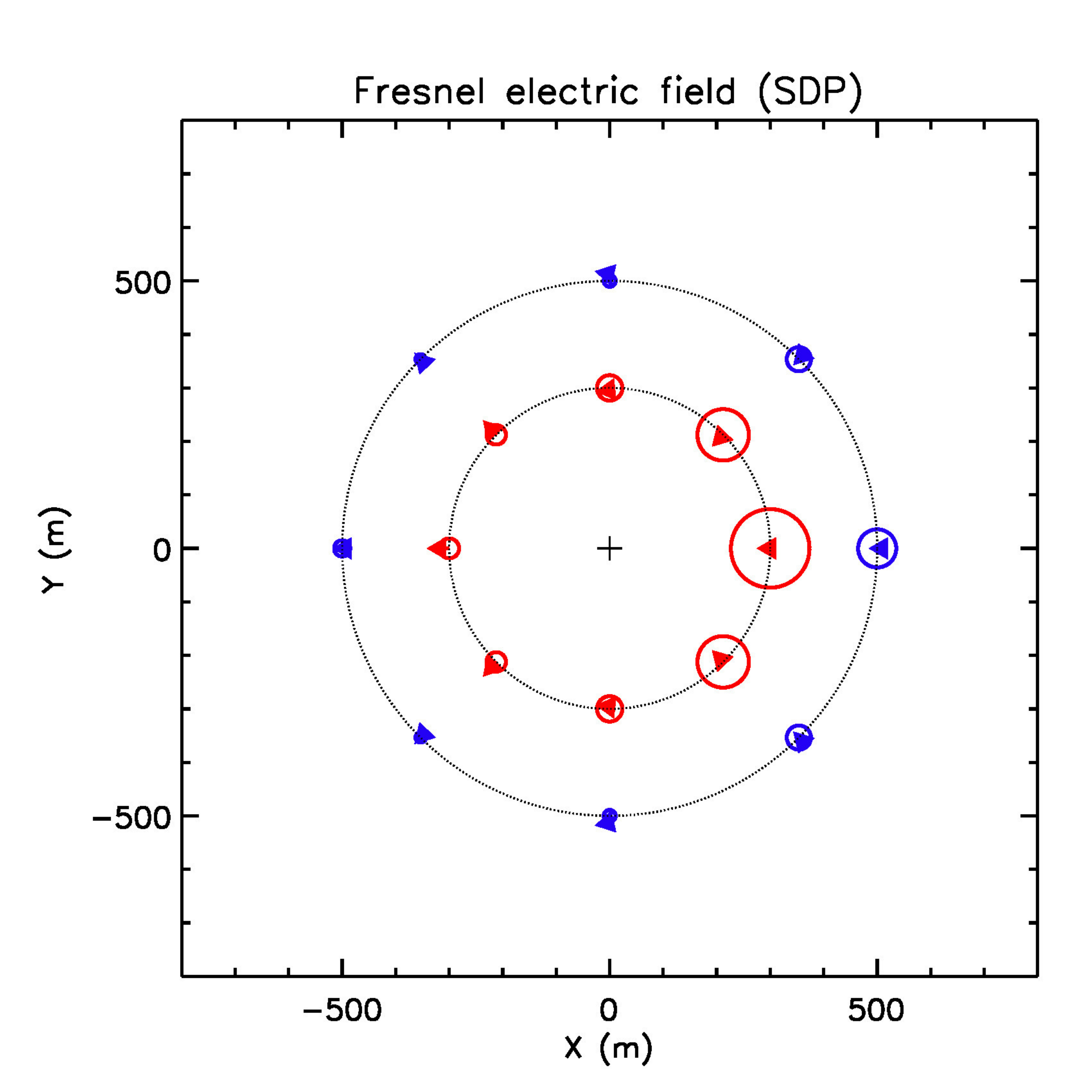}
\includegraphics[width=0.49\textwidth]{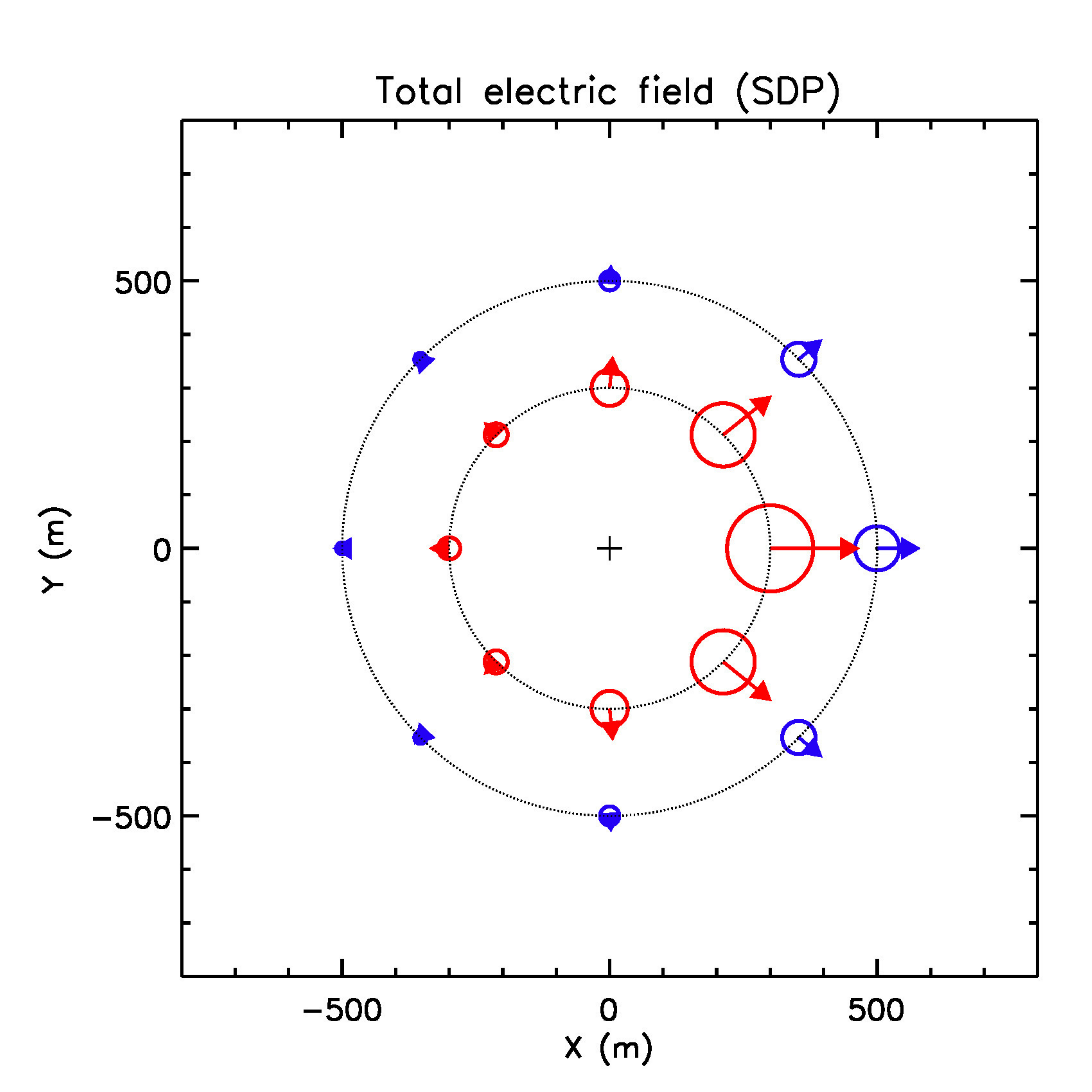}
\caption{Polarization map of the SDP emitted by a $30^\circ$ zenith angle
shower coming from the $-\hat y$ direction ($\varphi = 180^\circ$), with the core at $(0,0)$.
Black points indicate the position of the antennas, located on two rings $300$~m 
and $500$~m away from
the shower core. Arrows indicate the direction and amplitude of the horizontal polarization while
the radius of the circles indicate the amplitude of the vertical polarization (pointing always
on the $\hat z$ direction). 
Ground altitude is $1400$~m.
Top left: direct field. Top right: direct+reflected fields (Fresnel
approximation). Bottom: total exact field. See text for details.}
\label{fig:pol}
\end{figure}

When the shower is vertical, the SDP presents vertical polarization only. This is related to the
fact that, in our model, the deceleration at ground level lies along the $\hat z$ direction. 
To know if other polarizations are possible, we have simulated a $\theta = 30^\circ$ zenith 
angle shower coming from the azimuth $\varphi = 180^\circ$. We show a polarization map
for the direct, direct+reflected and total exact fields
in Fig.~\ref{fig:pol}, where the arrows indicate the horizontal polarization and the size of the circles
represent the vertical polarization. When comparing the direct vertical component 
(Fig.~\ref{fig:pol}, top left) with the exact one (Fig.~\ref{fig:pol}, bottom),
we see that the surface wave enhances the total vertical component. 
However, the horizontal polarization is slightly suppressed and its direction differs from
that obtained for the direct field.
We have checked that the Fresnel approximation (direct+reflected) does not suffice for understanding
this behavior, as depicted in Fig.~\ref{fig:pol} (top right). 
Our calculations show that the amplitude of the vertical
component of the SDP is similar to the direct field, and therefore the vertical polarization of the SDP
for a real shower should be similar to the ones in \cite{sdp}. However, the horizontal
(east-west and north-south) polarizations in \cite{sdp} for the SDP 
do not constitute a good approximation, in principle.

\begin{figure}
\includegraphics[width=0.7\textwidth]{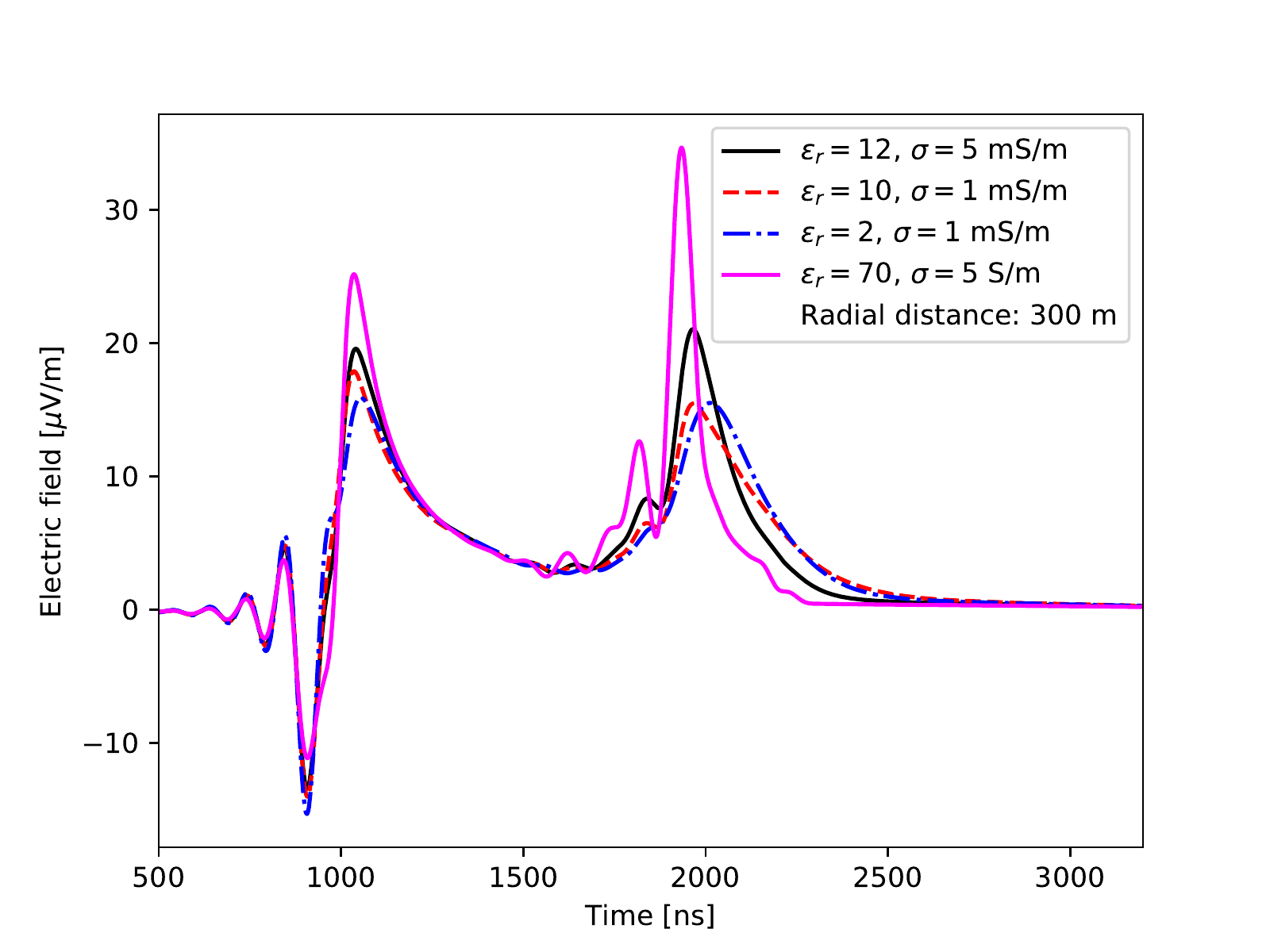}
\caption{Vertical component of the total exact electric field predicted by our shower model for
a vertical shower. Different grounds have been used: average soil at EXTASIS
(solid black line, $\epsilon_r = 12$, $\sigma = 5$~mS/m), damp AERA soil
(dashed red line, $\epsilon_r = 10$, $\sigma = 1$~mS/m), dry AERA soil
(dash-dotted blue line, $\epsilon_r = 2$, $\sigma = 1$~mS/m) and seawater
(solid magenta line, $\epsilon_r = 70$, $\sigma = 5$~S/m)
See text for details.}
\label{fig:soil}
\end{figure}

We show in Fig.~\ref{fig:soil} the influence of the type of ground on the total field. 
The ground altitude is fixed at $z_g = 0$~m. We have
chosen an average soil used for the EXTASIS antennas simulations ($\epsilon_r = 12$, 
$\sigma = 5$~mS/m), the soil present at the AERA experiment \cite{aerasoil} both for
dry ($\epsilon_r = 2$, $\sigma = 1$~mS/m) and damp conditions
($\epsilon_r = 10$, $\sigma = 1$~mS/m), and seawater \cite{itu}
($\epsilon_r = 70$, $\sigma = 5$~S/m). As it was already expected, since the emission from
the shower maximum can be approximated by a direct and a reflected component, and the reflected
component depends on the properties of the soil, the amplitude of the principal pulse (left) varies
with the type of soil used. In fact, choosing between a dry or damp ground changes the field more
than $10\%$, which reminds the importance of knowing the ground and its influence on the
measuring antennas. The SDP (right pulse) is also heavily affected by the type of ground, since
the properties of the surface wave are related to the properties of the ground. A higher
conductivity results in a larger surface wave, while the influence of the permittivity seems
to be more complicated.

Although the resulting SDP fields present an important component that is created by the surface wave,
which is a non-radiative type of field since it does not carry energy towards the infinity (it
vanishes at large distances from the interface, where the radiative direct and reflected components
dominate), we can see in Fig.~\ref{fig:disttime}, top, that the SDP amplitudes for a vertical shower
according to the model fall with the inverse of the distance to the shower core, as if it were a pure
radiation field. This result is in agreement with the direct field calculation in \cite{sdp}.
We retrieve as well that the SDP amplitude increases when the ground altitude is higher, due to the
larger number of particles arriving at the ground.

We show in Fig.~\ref{fig:disttime}, bottom, the arrival time of the SDP as a function of distance. 
$t=0$ is the time the shower reaches ground. The data for 1400 and 2500~m of altitude have been 
offset for clarity. The arrival times grow linearly with distance as evidenced by the superposed linear fits.
The inverses of the fit slopes are $0.2966$, $0.2961$ and $0.2958$~m$\,$ns${}^{-1}$ for
0, 1400, and 2500~m of altitude respectively. The speed of light in air is $\sim 0.2998$~m$\,$ns${}^{-1}$,
which is slightly larger, reflecting the fact that the surface wave travels close to the boundary
and therefore has a lower effective speed than the direct and reflected components. However,
there is a linear relation between the SDP arrival time computed using the surface wave and the
distance to the shower core, as it was the case for the direct field only \cite{sdp}.

\section{Summary and conclusions}

Since the field of a dipole can be used to calculate the field created by more complex configurations,
we have discussed the exact frequency-domain electric field created by a unit dipole in two
semi-infinite media separated by a planar boundary. If the dipole and the observer are in the
same medium, the field can be separated in three fields: a direct field, an ideal image field, and
an integral that contains information about the lateral wave created by the boundary. If
the observer and the dipole lie in different media, the whole field can be expressed with a single
integral that yields the field that passes through the boundary. In both cases, the integrals
have to be numerically evaluated, in general. The integrals contain Bessel functions with long
oscillating tails that present a very low convergence. To keep the computation time reasonable,
we have chosen the partition extrapolation method as our method of integration.

Then, we have presented a frequency-domain equation produced by a particle track that takes into account
the effect of the boundary on the electric field at all frequencies. This field has been obtained by
integrating the solution for the electric field of a dipole, previously obtained.
As with the dipole case, the track field can be divided into direct, image, and integral fields if the
observer and the track lie in the same medium (Eq.~\eqref{eq:etracksum}). 
However, in this case the three partial fields have
to be numerically integrated, in general. We have shown that the Fraunhofer approximation for
integrating the direct and image fields is a good approximation. In the far field, where either the
track or the observer are far away from the boundary, the track field is expressable by the sum of
a direct and reflected field (using Fresnel coefficients). Our formula is also equivalent to 
the ZHS-TR formalism \cite{motloch} in the far field. We have also obtained a solution for the case
when the observer and the track are in different media (Eq.~\eqref{eq:etracksumunder}), which
reduces in the far field to a transmitted wave calculated using a modified Fresnel coefficient as
in \cite{endpoints,motloch}. The transmitted field calculated in the present work is also consistent 
with ZHS-TR. As a rule of thumb, when either the distance from track to boundary or the distance
from observer to boundary is $3$ times greater than the observation wavelength, the relative
error of the far-field approximation is less than $~10\%$.

We have studied the spectra for a track in air near a soil ground. For frequencies under
10~MHz, which is the domain of the EXTASIS experiment, the boundary creates a surface wave that
interferes constructively with the direct field. The behavior of these spectra with radial distance shows
that, under 10~MHz, the contribution of the surface wave is quite relevant up to radial distances
of several hundreds of meters. We have also calculated the field from a short underground track seen
by an observer in the atmosphere and found that the underground field is two orders of magnitude
than the field emitted by a typical particle track traveling in air near the boundary, which implies
that the underground particles can be ignored.

Since the computation time for the exact electric field from a realistic shower 
simulated with a Monte Carlo code seems unmanageable,
we have proposed a simple model for a EAS. We have calculated the exact field under
10~MHz for a set of
one-dimensional subshowers presenting a Gaisser-Hillas longitudinal profile together
with a NKG lateral distribution. The results show
that the emission coming from the shower maximum is not modified by the boundary field, but
when the shower is near the ground the influence of the boundary is rather important. The
exact calculation confirms that a sudden death pulse (SDP) is created when the shower
abruptly stops at ground level, which had already been obtained in a less rigorous way in
\cite{sdp} using the direct field only. The order of magnitude of the vertical
component of the SDP
predicted by the exact calculation is the same as that predicted by the exact calculation, which
means that the direct emission is a decent approximation for the vertical field and the
calculations of \cite{sdp} for the vertical component remain valid. The horizontal
components, on the other hand, are not well approximated by the direct emission or
the Fresnel approximation (direct+reflected emission), and the surface wave is needed to
produce an accurate amplitude and polarization.

The exact calculation also shows that, although the surface wave is a non-radiative field, the
exact SDP amplitude falls with the inverse of the distance to the shower core, as pure radiation
field would do. The arrival time of the SDP maximum is proportional to the distance to the shower
core as well, with a velocity of propagation speed slightly inferior to the speed of light in air,
since the surface wave propagates near the boundary and not through the direct path joining
emitter and observer.

We must point out that while the present work provides a way of calculating the electric field of a track
when a planar boundary is present, we have not discussed the voltage this field would induce
in an antenna. Antennas are well understood when working in the far-field regime, with radiating
sources. When the sources are near and the field is a mixture of radiating and
non-radiating components the reception patterns become more complicated. A proper understanding
of the antenna response in the near-field regime of an EAS is capital for a correct prediction of
the final voltage.

\begin{acknowledgments}
We thank the R\'{e}gion Pays de la Loire for its financial support of the Groupe Astro
of Subatech and in particular for its contribution to the EXTASIS experiment.
\end{acknowledgments}

\appendix

\section{Obtaining the horizontal direct field from the vertical field}
\label{sec:vtoh}

The direct field created by the horizontal dipole can be obtained from the direct
field emitted by the vertical dipole with the appropriate coordinate transformations.
This stems from the fact that the direct field is precisely the field calculated as
if there were no boundary. Starting with Eq.~\eqref{eq:Evd}, let us make $z' = 0$
and place an observer at $(x,0,z)$, so that $E_\rho = E_x$ and $E_y = 0$, without loss of generality.
Then, we make the following change of coordinates - we rotate $-\pi$ along the y axis, 
so that the dipole is oriented towards the $+\hat x$ unit vector in the new frame:
\begin{eqnarray}
x' = z \nonumber \\
y' = y \nonumber \\
z' = -x.
\label{eq:coordtrans1}
\end{eqnarray}
This transformation (note that in this case $z'$ denotes the new $z$ coordinate and not
the dipole height) must be applied to the electric field as well:
\begin{eqnarray}
E_{x'} = E_z \nonumber \\
E_{y'} = E_y \nonumber \\
E_{z'} = -E_x
\label{eq:vectrans1}
\end{eqnarray}
With the observer at $(x,0,z)$, Eq.~\eqref{eq:Evd} changes to the form:
\begin{eqnarray}
E^{d}_{x} & = & -\frac{\omega\mu_0}{4\pi k_1^2} e^{i k_1 r_1}
\left( \frac{ik_1^2}{r_1} - \frac{3k_1}{r_1^2} - \frac{3i}{r_1^3}  \right)
\left( \frac{x}{r_1} \right) \left( \frac{z}{r_1} \right) \nonumber \\
E^{d}_{z} & =  &\frac{\omega \mu_0}{4\pi k_1^2} e^{i k_1 r_1}
\left[ \frac{ik_1^2}{r_1} - \frac{k_1}{r_1^2} - \frac{i}{r_1^3}
- \left( \frac{z}{r_1} \right)^2 \left( \frac{ik_1^2}{r_1} - \frac{3k_1}{r_1^2}
- \frac{3i}{r_1^3} \right) \right]
\end{eqnarray}
Applying Eqs.~\eqref{eq:coordtrans1} and \eqref{eq:vectrans1}, along with the
identity $x'^2 = r_1^2-z'^2$, we arrive at
\begin{eqnarray}
E^{d}_{x'} & = & \frac{\omega\mu_0}{4\pi k_1^2}
e^{ik_1 r_1} \left[ \frac{2k_1}{r_1^2} + \frac{2i}{r_1^3} + 
\frac{(z')^2}{r_1^2} \left( \frac{i k_1^2}{r_1} - \frac{3k_1}{r_1^2}
- \frac{3i}{r_1^3} \right) \right]
 \nonumber \\
E^{d}_{z'} & =  &-\frac{i\omega \mu_0}{4\pi k_1^2} e^{i k_1 r_1}
\left( \frac{x'}{r_1} \right) \left( \frac{z'}{r_1} \right)
\left( \frac{k_1^2}{r_1} + \frac{3ik_1}{r_1^2}
- \frac{3}{r_1^3} \right),
\end{eqnarray}
which is the particular case of Eq.~\eqref{eq:Ehd} when the observer is located at $(x,0,z)$
and the dipole is at the origin. We now perform a rotation of an angle $\alpha$ around the 
$x'$ axis. Using the fact that $y' = 0$ and $E_{y'} = 0$, the transformations are:
\begin{eqnarray}
x'' = x' \nonumber \\
y'' = \sin\alpha \ z' \nonumber \\
z'' = \cos\alpha \ z'
\label{eq:coordtrans2}
\end{eqnarray}
and
\begin{eqnarray}
E_{x''} = E_{x'} \nonumber \\
E_{y''} = \sin\alpha E_{z'} \nonumber \\
E_{z''} = \cos\alpha E_{z'}.
\label{eq:vectrans2}
\end{eqnarray}
Let us begin with $E_{z''}$. Eqs.~\eqref{eq:coordtrans2} and \eqref{eq:vectrans2} imply that
\begin{equation}
E^{d}_{z''} = \cos\alpha E^{d}_{z'} =
- \cos\alpha \frac{i\omega \mu_0}{4\pi k_1^2} \ e^{i k_1 r_1}
\left( \frac{x''}{r_1} \right) \left( \frac{z''}{\cos\alpha \  r_1} \right)
\left( \frac{k_1^2}{r_1} + \frac{3ik_1}{r_1^2}
- \frac{3}{r_1^3} \right),
\end{equation}
and using cylindrical coordinates $(\rho,\varphi,z'')$ we have that $x'' = \rho\cos\varphi$ and
therefore:
\begin{equation}
E^{d}_{z''} = -\frac{i\omega \mu_0}{4\pi k_1^2} \ e^{i k_1 r_1}  \cos\varphi
\left( \frac{\rho}{r_1} \right) \left( \frac{z''}{r_1} \right)
\left( \frac{k_1^2}{r_1} + \frac{3ik_1}{r_1^2}
- \frac{3}{r_1^3} \right),
\label{eq:ezend}
\end{equation}
which is the same as the $z$ component in Eq.~\eqref{eq:Ehd} if we drop the two primes
from the $z''$. On the other hand, $E_{x''}$ and $E_{y''}$ can be written as:
\begin{equation}
E^{d}_{x''} = E^{d}_{x'} = 
\frac{\omega\mu_0}{4\pi k_1^2}
e^{ik_1 r_1} \left[ \frac{2k_1}{r_1^2} + \frac{2i}{r_1^3} + 
\frac{(z'')^2}{\cos^2\alpha \ r_1^2} \left( \frac{i k_1^2}{r_1} - \frac{3k_1}{r_1^2}
- \frac{3i}{r_1^3} \right) \right]
\label{eq:exmiddle}
\end{equation}
\begin{equation}
E^{d}_{y''} = \sin\alpha E^{d}_{z'} = -\frac{\sin\alpha}{\cos\alpha} 
\frac{i\omega \mu_0}{4\pi k_1^2} e^{i k_1 r_1}
\left( \frac{\rho\cos\varphi}{r_1} \right) \left( \frac{z''}{r_1} \right)
\left( \frac{k_1^2}{r_1} + \frac{3ik_1}{r_1^2}
- \frac{3}{r_1^3} \right).
\label{eq:eymiddle}
\end{equation}
The sinus and cosinus of $\varphi$ can be expressed in the following way:
\begin{equation}
\cos\varphi = \frac{x''}{\rho}
\end{equation}
\begin{equation}
\sin\varphi = \frac{y''}{\rho} = \frac{\sin\alpha \ z'}{\rho} = 
\frac{\sin\alpha \ z''}{\cos\alpha \ \rho},
\end{equation}
from which we derivate the identity:
\begin{equation}
\sin\varphi \frac{\sin\alpha}{\cos\alpha} z'' = \frac{\sin^2\alpha}{\rho} \frac{{z''}^2}{\cos^2\alpha}
\label{eq:sinphiid}
\end{equation}
We know that the radial field can be obtained by combining the horizontal fields:
\begin{equation}
E_\rho = E_{x''}\cos\varphi + E_{y''}\sin\varphi,
\end{equation}
which, in conjunction with Eqs.~\eqref{eq:exmiddle}, ~\eqref{eq:eymiddle} and \eqref{eq:sinphiid}
gives:
\begin{equation}
E^d_\rho = 
\frac{\omega\mu_0}{4\pi k_1^2} \cos\varphi
e^{ik_1 r_1} \left[ \frac{2k_1}{r_1^2} + \frac{2i}{r_1^3} + 
\frac{(z'')^2}{r_1^2} \left( \frac{i k_1^2}{r_1} - \frac{3k_1}{r_1^2}
- \frac{3i}{r_1^3} \right) \right],
\label{eq:erhoend}
\end{equation}
which is consistent with Eq.~\eqref{eq:Ehd}. Finally, expressing the azimuthal field as
\begin{equation}
E_\phi = -E_{x''}\sin\varphi + E_{y''}\cos\varphi,
\end{equation}
along with the following identity:
\begin{equation}
\frac{\sin\varphi \ {z''}^2}{r_1^2\cos^2\alpha} + \cos^2\varphi \frac{\sin\alpha}{\cos\alpha}
\frac{\rho z''}{r_1^2} = \sin\varphi \frac{z''^2}{r_1^2} + \frac{\sin\alpha \rho z'}{r_1^2}
= \frac{\sin\varphi}{r_1^2}(\rho^2 + {z''}^2) = \sin\varphi,
\end{equation}
leads us to the final expression for the azimuthal field:
\begin{equation}
E^d_\varphi = -\frac{\omega \mu_0}{4\pi k_1^2} e^{ik_1r_1} \sin\varphi
\left[ \frac{ik_1^2}{r_1} - \frac{k_1}{r_1} - \frac{i}{r_1^3} \right],
\label{eq:ephiend}
\end{equation}
which is the same as in Eq.~\eqref{eq:Ehd}. Eqs.~\eqref{eq:ezend}, \eqref{eq:erhoend} and
\eqref{eq:ephiend} show that the direct field of a horizontal dipole can be obtained from
the field of a vertical dipole.

\bibliography{surface_wave}

\end{document}